\documentclass[aps,prx,twocolumn,superscriptaddress, longbibliography]{revtex4-2}
\usepackage{amsmath}    
\usepackage{graphicx}
\usepackage{enumitem}
\usepackage{bm}
\UseRawInputEncoding
\usepackage{booktabs}
\usepackage{bbm}
\usepackage{xcolor,soul}
\usepackage{amssymb}
\usepackage{mathdots}
\definecolor{darkblue}{HTML}{3771C8}
\usepackage[colorlinks=true,
    pdfborder={0 0 0},
    linkcolor=darkblue,
    citecolor = darkblue, 
    urlcolor = darkblue
]{hyperref}
\usepackage[capitalize,nameinlink]{cleveref}
\usepackage[normalem]{ulem}
\usepackage{braket}

\renewcommand{\selectlanguage}[1]{}
\begin{document}
\title{Many-body quantum optics in a cascaded chiral network}

\author{Frank Yang}
    \email[These authors contributed equally.]{}
    \affiliation{Moore Laboratory of Engineering, California Institute of Technology, Pasadena, California 91125}
    \affiliation{Institute for Quantum Information and Matter, California Institute of Technology, Pasadena, California 91125}
    
\author{Parth S. Shah}
    \email[These authors contributed equally.]{}
    \affiliation{Moore Laboratory of Engineering, California Institute of Technology, Pasadena, California 91125}
    \affiliation{Institute for Quantum Information and Matter, California Institute of Technology, Pasadena, California 91125}
    
\author{Chaitali Joshi}
    \affiliation{Moore Laboratory of Engineering, California Institute of Technology, Pasadena, California 91125}
    \affiliation{Institute for Quantum Information and Matter, California Institute of Technology, Pasadena, California 91125}
    
\author{Mohammad Mirhosseini}
    \email[mohmir@caltech.edu; http://qubit.caltech.edu]{}
    \affiliation{Moore Laboratory of Engineering, California Institute of Technology, Pasadena, California 91125}
    \affiliation{Institute for Quantum Information and Matter, California Institute of Technology, Pasadena, California 91125}

\date{\today} 

\begin{abstract} 

Chiral quantum emitters interact with light only in one propagation direction, allowing them to be linked into cascaded systems in which photons mediate ordered, long-range interactions. Such systems are predicted to host novel regimes of many-body physics of light and matter. Exploring these regimes requires arrays of identical quantum emitters with directional, low-loss coupling to guided photons, a combination that has thus far remained experimentally out of reach. Here we realize a cascaded network of superconducting qubits using an architecture that overcomes these bottlenecks. We implement a four-qubit chain spanning two modules, with separations ranging from millimeters to half a meter, and exploit the shared waveguide as a dissipative resource to stabilize reconfigurable entanglement, reaching a genuinely multipartite regime unavailable in reciprocal baths. By scattering weak pulses off the chain, we observe photons sorted in time by photon number, a signature of the strong photon-photon interactions mediated by the emitters. Together, these results provide experimental access to many-body light-matter regimes that are beyond the reach of reciprocal systems.

\end{abstract}

\maketitle

\section*{Introduction}

\begin{figure*}[t!]
\centering
\includegraphics[width=1\linewidth]{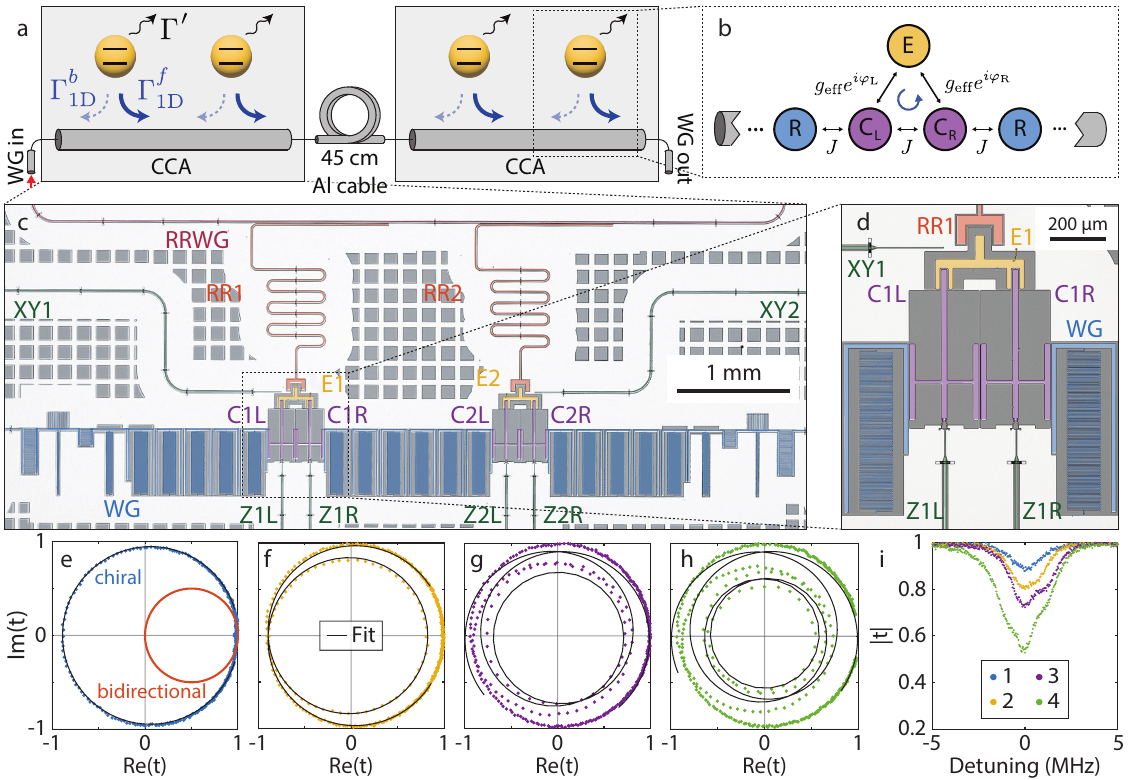}
\caption{\textbf{Cascaded artificial atoms.} (a) Schematic of cascaded chiral qubits coupled to a waveguide. The four-qubit chain is comprised of two device chips connected by a superconducting cable. (b) Diagram of qubit-waveguide coupling, showing an emitter E coupled to two adjacent couplers C with phases $\varphi_\mathrm{L}$ and $\varphi_\mathrm{R}.$ Couplers are embedded in a chain of coupled resonators R which comprise the waveguide. (c) Optical image of a device chip, which contains two emitter qubits (yellow; E1, E2), couplers (purple; C1L, C1R, C2L, C2R), and CCA waveguide (blue, WG). Each coupler has a dedicated flux line (green, Z1L, Z1R, Z2L, Z2R), and each emitter qubit is equipped with an XY line (green, XY1, XY2) and readout resonator (orange, RR1, RR2). Readout resonators are probed through a separate waveguide (pink, RRWG). (d) Zoomed-in image of emitter and couplers embedded in the CCA. (e,f,g,h) Measured complex transmission of (e) one, (f) two, (g) three, (h), and four qubits, showing a phase accumulation of $2\pi N$ for $N$ qubits. Also shown in (e) is complex transmission (red) of an ideal bidirectional qubit, which does not cross the origin. (i) Amplitude of measured transmission for one, two, three and four cascaded qubits.}
\label{fig:Fig1}
\end{figure*}

Chirality -- the breaking of mirror symmetry -- is ubiquitous across the physical and life sciences, with consequences ranging from the action of modern pharmaceuticals \cite{brown1989chemical} to the structure of the fundamental forces \cite{wu1957experimental}. Directional light-matter interaction is in fact built into even the earliest microscopic models of optics \cite{bohren2008absorption}. As described by Lorentz, a wave propagating through a dense medium drives the bound electrons of each atom, and the re-radiated fields interfere in the forward direction to set the medium's refractive index, while backward radiation cancels. While this directionality arises for free from the collective response of a dense ensemble, it is at the level of the single atom that chirality becomes a powerful resource for quantum optics.

Coupling individually chiral quantum emitters to a shared one-dimensional bath results in a cascaded quantum system, in which photons mediate ordered, one-way interactions between distant atoms \cite{10.1103/physrevlett.70.2273, 10.1103/physrevlett.70.2269}. Such systems are predicted to open many-body regimes inaccessible to reciprocal ones. On the light side, the non-reciprocal scattering produces correlated multi-photon transport and photon bound states \cite{mahmoodian2020dynamics, morvan2022formation}; on the matter side, collective decay into the shared bath can stabilize entangled and topologically ordered states of the emitters \cite{stannigel2012driven, pichler2015quantum, barik2018topological}. The same one-way coupling also realizes a quantum network with all-to-all connectivity between nodes \cite{cirac1997quantum, mcintyre2025protocols, 10.1103/physrevlett.118.133601}, with applications spanning quantum error correction \cite{yoder2025tour} and dissipative quantum computation \cite{verstraete2009quantum, mi2024stable}.

Reaching these regimes, however, places stringent demands on the emitter-bath interface. Each quantum emitter must couple exclusively to one propagation direction while suppressing all other decay channels, so that a photon released by one emitter can be reabsorbed by the next with near-unit probability \cite{Lodahl2017Jan}. Meeting these requirements for a single emitter has proven challenging, while extending them to many emitters without compromising coherence, directionality, or tunability has remained the central obstacle to realizing many-body cascaded systems.

At optical frequencies, chirality arises from spin-momentum locking of guided modes \cite{le2015nanophotonic}. Existing implementations have demonstrated either deterministic interfaces -- using quantum dots \cite{sollner2015deterministic}, or large cascaded arrays -- using trapped atoms \cite{liedl2024observation}, but not both simultaneously because of frequency disorder and probabilistic coupling, respectively. Microwave implementations have achieved deterministic chiral interfaces using ferrite circulators \cite{reuer2022realization} or waveguide interference \cite{kannan2023demand, joshi2023resonance}, yet remain limited to one or two emitters by loss, frequency crowding, and bulky off-chip components \cite{almanakly2025deterministic, irfan2025autonomous}. Moreover, composite-emitter interfaces lose their directionality beyond the single-photon regime \cite{kannan2023demand}, preventing access to strongly nonlinear many-body dynamics. Consequently, deterministic cascaded arrays have yet to be demonstrated experimentally.

Here we realize the smallest deterministic cascaded network in which qualitatively new many-body physics arises, comprising four superconducting qubits coupled to a shared microwave waveguide. Our architecture combines dispersion engineering and on-chip parametric generation of chiral coupling to overcome the loss, frequency-crowding, and nonlinearity limitations of previous implementations, providing strong, tunable, and unidirectional emitter-waveguide coupling. We use these capabilities to access complementary many-body regimes in both matter and light. By exploiting the dark states of the cascaded qubit chain that are formed by collective decay into the waveguide, we stabilize reconfigurable four-qubit entanglement, generating both dimerized states and genuine multipartite entanglement that is not accessible in reciprocal systems. When driving the chain with weak coherent pulses, we observe scattered light sorted in time into bound states of well-defined photon number, a direct manifestation of the strong photon-photon interactions mediated by the emitters. Combined, these results bring the many-body light-matter phenomena predicted for cascaded quantum systems into experimental reach.

\section*{Device overview}

The cascaded system is comprised of four `emitter' qubits parametrically coupled to a microwave waveguide (\cref{fig:Fig1}a). Each qubit decays to the waveguide in the forward direction with emission rate $\Gamma^{f}_{\mathrm{1D}}$. Meanwhile, un-desired decoherence arises from emission to the backward waveguide direction ($\Gamma^{b}_{\mathrm{1D}}$) and the intrinsic qubit decoherence ($\Gamma'$). The Purcell factor, $P_\mathrm{1D} = \frac{\Gamma^f_{\mathrm{1D}}}{\Gamma^b_{\mathrm{1D}}+\Gamma'}$, describes the competition between these rates and quantifies the efficiency of a chiral qubit. The waveguide consists of a coupled cavity array (CCA), which can be described by a tight-binding Hamiltonian with nearest-neighbor couplings (see \cref{fig:Fig1}b). The emitter interacts with the CCA via coupling to two adjacent array elements, referred to as couplers. By applying a parametric drive to modulate the coupler frequencies, a phase can be imparted to the emitter-waveguide couplings, $g_\mathrm{eff}e^{i\varphi_\mathrm{L}}$ and $g_\mathrm{eff}e^{i\varphi_\mathrm{R}}$. Tuning the relative coupling phases $\varphi_\mathrm{L}$ and $\varphi_\mathrm{R}$ creates a synthetic gauge field in the emitter-coupler-coupler loop \cite{clerk2022introduction}, which breaks the time-reversal symmetry and produces a chiral qubit-waveguide interaction \cite{joshi2023resonance} (See \cref{appendix:Design:chiral}).

We construct a four-qubit chain by connecting two device chips with a superconducting aluminum cable, as shown in \cref{fig:Fig1}a. Each chip contains two emitter qubits. A detailed view of a device chip is given in \cref{fig:Fig1}c, showing the emitter qubits, couplers, and coupled-resonator waveguide. \cref{fig:Fig1}d shows a single emission module, comprising a fixed-frequency transmon emitter and two flux-tunable transmon couplers. Microwave tones applied to the coupler flux lines generate the requisite coupling phases, up-converting the emitter frequency via three-wave mixing (see \cref{appendix:Design:CMT}). 

Our system is designed to overcome the two limitations that have, until now, prevented access to the quantum many-body physics of deterministic chiral emitters. The first concerns the non-linearity of the emitter. ``Giant-molecule" interfaces, formed from qubit pairs with distributed waveguide coupling \cite{kannan2023demand}, exhibit a chiral response only within the single-excitation manifold, so they lose directionality under strong driving and cannot support the non-linear dynamics required for driven-dissipative entanglement generation or photon sorting. Our emitters are instead ``giant atoms," individual qubits coupled to the waveguide at two points \cite{joshi2023resonance}, which retain their full quantum non-linearity and enable the strongly-driven, many-body regimes we explore here \cite{stannigel2012driven, pichler2015quantum}. The second limitation concerns scaling. Giant-atom interfaces have so far been restricted to a single emitter by frequency crowding, since the resonant filter elements used to suppress spurious sidebands occupy spectral resources that cannot be replicated for many qubits \cite{joshi2023resonance}. By instead embedding tunable couplers directly within the coupled cavity array, we use the array's bandgap to suppress these parasitic sidebands without dedicated filters \cite{joshi2023resonance, almanakly2025deterministic}. This removes the crowding bottleneck and simultaneously improves the quality of each interface (see \cref{appendix:Design:CMT}). Design, fabrication, and device details are provided in \cref{appendix:Design:circuit}, \cref{appendix:Methods:fab}, and \cref{appendix:Device:summary}.

We confirm the chiral strong coupling of each emitter using elastic scattering, coupling one qubit at a time to the waveguide and measuring the transmission of a weak drive (\cref{appendix:Methods:setup}). Each emitter exhibits a $2\pi$ phase response across its linewidth and a $\pi$ phase shift on resonance, the signature of unidirectional strong coupling \cite{joshi2023resonance}, with directionality ratios up to $\eta_\mathrm{d} = \Gamma_\mathrm{1D}^f/\Gamma_\mathrm{1D}^b \sim 100$ and Purcell factors up to $P_\mathrm{1D}\sim20$ (\cref{fig:Fig1}e). This effectively eliminates the parametric-drive-induced decoherence that limited previous work \cite{joshi2023resonance}, shifting the limiting factors to intrinsic qubit quality and waveguide thermalization (\cref{appendix:Device:chirality tuneup}). Subsequently, we test the defining property of the network by cascading resonant emitters. In an ideal cascaded system, the absence of information backflow implies that the total weak-power transmission of $N$ chiral qubits is the product of the individual transmissions, $t = \prod_{j=1}^N t_j$. The measured transmission follows this expectation, where the complex coefficient makes $N$ revolutions around the origin as emitters are added (see \cref{fig:Fig1}f-i), in agreement with a transmission-matrix model (see \cref{appendix:Modeling:ABCD}).

\section*{Sub-radiant states and  driven-dissipative entanglement stabilization}

\begin{figure*}[t!]
\centering
\includegraphics[width=1\linewidth]{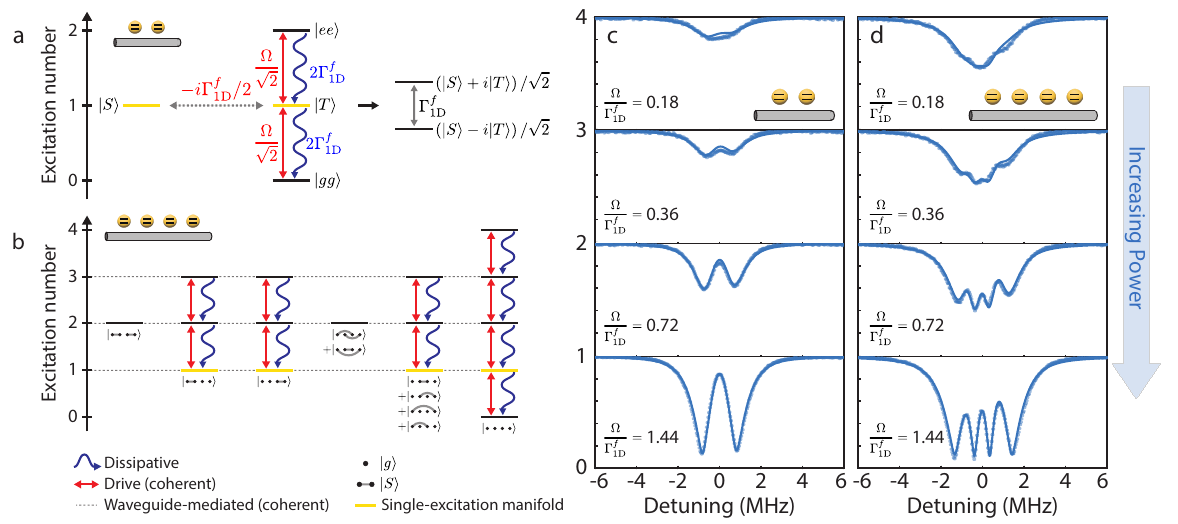}
\caption{\textbf{Collective sub-radiant states.} (a, b) Level structure for (a) two and (b) four cascaded qubits \cite{pichler2015quantum}. For two qubits, $|S\rangle=\frac{|ge\rangle - |eg\rangle}{\sqrt{2}}$ and $|T\rangle=\frac{|ge\rangle + |eg\rangle}{\sqrt{2}}$. In each case, the Hilbert space contains super-radiant and sub-radiant manifolds coupled coherently by waveguide-mediated interactions. The resulting hybridization (shown in (a) for two qubits) can be observed by power-dependent elastic scattering, where the (c) two- and (d) four-qubit transmission splits in two and four dips, respectively. $\Omega$ is the Rabi drive amplitude. Measured data (dots) is fit using master equation simulations (solid lines; see \cref{appendix:Modeling:SLH ME}, \cref{appendix:Modeling:power dep transmission}).}
\label{fig:Fig2}
\end{figure*}

Under a continuous classical drive, the collective decay into the shared waveguide bath can be used as a resource to steer the qubits toward an entangled steady state \cite{shah2024stabilizing, irfan2025autonomous, andres2025entangling, Almanakly2026drivendissipative}. This mechanism is known as driven-dissipative stabilization. It prepares long-lived entanglement on demand, requiring only a shared reservoir and a drive rather than pulse-level control. While stabilization through a shared waveguide does not require directionality in general \cite{shah2024stabilizing}, chirality changes what can be achieved in this setting. The absence of information backflow lifts the phase-length sensitivity that constrains the practical range of bidirectional schemes and grants access to pure multipartite entangled steady states beyond the reach of reciprocal baths \cite{stannigel2012driven, pichler2015quantum}. We exploit both of these features below.

We first study the dark states of the cascaded qubit chain. The collective sub-radiant subspace of $N$ qubits can be found by adding each qubit -- an effective spin-1/2 -- to a collective angular momentum. This separates collective states into distinct spin manifolds, as shown in \cref{fig:Fig2}a and b for two and four qubits, respectively. In each spin manifold, we refer to the lowest-excitation-number state as sub-radiant because it has no direct waveguide decay. Entangled sub-radiant states instead interact with the waveguide via chiral waveguide-mediated couplings to super-radiant states. These interactions are directly evident in the power-dependent elastic scattering. By increasing the drive power for $N$ qubits, we observe a splitting in the transmission into $N$ dips, which arises from the coupling between $N-1$ single-excitation sub-radiant states and a single super-radiant state. Measurements are shown in \cref{fig:Fig2}c,d for $N=2$ and $N=4$ qubits. Transmission dips become deeper at higher drive powers because of power broadening, which arises from the cascaded non-linearity. Also, the single-excitation manifold dominates here because higher excited levels are negligibly populated for probe powers below full qubit saturation. 

\begin{figure*}[t!]
\centering
\includegraphics[width=1\linewidth]{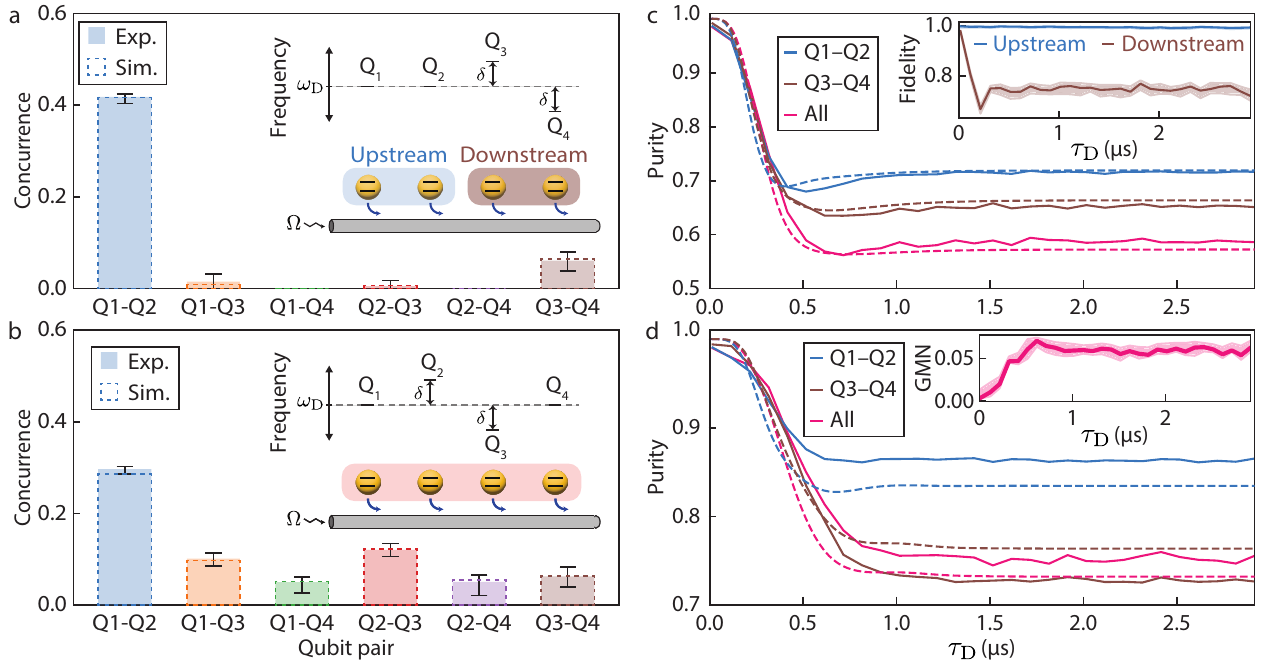}
\caption{\textbf{Entanglement stabilization.} The qubit chain is driven via the waveguide for variable time $\tau_\mathrm{D}$, after which qubit-waveguide couplings are turned off, and the state is measured (see \cref{appendix:Methods:state tomography}, \cref{appendix:Methods:ampnoise}). Measured data are fit to master equation simulations, shown in dotted lines (see \cref{appendix:Modeling:ddre fits}). (a) Measured pair-wise concurrences showing dimerization, as evidenced by finite concurrence between qubits 1 and 2 and qubits 3 and 4, with all other concurrences $\approx0$. Inset: Dimerization schematic and detunings: $\delta_1 = -\delta_2 = 0$ MHz, $\delta_3 = -\delta_4 = 2$ MHz. (b) Measured purity of the four-qubit state and two dimers vs. time, showing preparation of upstream dimers prior to downstream dimers. Inset: fidelity between upstream (downstream) dimers measured in (1) the presence of the downstream (upstream) dimer and (2) the absence of the downstream (upstream) dimer. (c) Measured pair-wise concurrences for a multipartite entangled state, showing finite entanglement between all qubit pairs. Inset: Multipartite entanglement schematic and detunings: $\delta_1 = -\delta_4 = 0$ MHz, $\delta_2 = -\delta_3 = 2$ MHz. (d) Measured purity vs. time of the multipartite entangled state. Inset: Genuine multipartite negativity vs. time, showing stabilization.}
\label{fig:ddre}
\end{figure*}

The simplest sub-radiant state to prepare is a two-qubit dimer $\approx|S\rangle$, where the emission from an upstream qubit is perfectly absorbed by a downstream qubit, decoupling the pair from the coherent dynamics \cite{stannigel2012driven}. For a four-qubit chain, setting the qubit-drive detunings to $\delta_1 = -\delta_2$, $\delta_3 = -\delta_4$ (as shown in \cref{fig:ddre}a) results in dimerization, where the approximate steady state is given by $|\Psi\rangle \approx |S\rangle_{12} |S\rangle_{34}$. We measure the entanglement structure of this steady state in \cref{fig:ddre}a, which shows the concurrence between different qubit pairs. We observe finite concurrence between qubits 1 and 2 and qubits 3 and 4, while all other pairs exhibit no detectable entanglement, confirming the dimerization. Stabilization dynamics are shown in \cref{fig:ddre}b, which plots the purity of the four-qubit chain and two dimers. The purity of the upstream dimer (qubits 1 and 2) stabilizes to a steady-state value before the downstream dimer (qubits 3 and 4), reflecting the directionality of the chain. This one-way structure is what makes driven-dissipative entanglement generation practical over long distances. In a bidirectional waveguide, stabilization succeeds only at specific phase-length separations \cite{shah2024stabilizing}, whereas the absence of backflow removes that constraint and allows the same protocol to operate across emitters separated from millimeters on a chip to nearly half a meter over a cable.

We highlight the absence of information backflow in the qubit chain by repeating the state generation experiment for three cases: (1) four qubits coupled to the waveguide, (2) only upstream dimer coupled to the waveguide, and (3) only downstream dimer coupled to the waveguide. We find that, in (1) and (2), the upstream dimer stabilizes to the same steady state, with a fidelity $\approx 1$ between these cases, as shown in \cref{fig:ddre}b inset. In other words, the presence of the downstream dimer does not affect the dynamics of the upstream dimer. In contrast, the fidelity of the downstream dimer between (1) and (3) does not reach $\approx 1$ because the upstream dimer strongly affects the downstream dynamics. 

Re-arranging the qubit detunings to $\delta_1 = -\delta_4$ and $\delta_2 = -\delta_3$ is expected to result in multipartite entanglement in a state that ideally approaches a valence bond state \cite{pichler2015quantum}. With these detunings, we measure pair-wise concurrences shown in \cref{fig:ddre}c, where, in contrast to the dimerized case, all concurrences are finite, reflecting a change in the entanglement structure. The system dynamics are shown in \cref{fig:ddre}d, again showing stabilization of upstream qubits prior to downstream qubits. We confirm the genuine multipartite entanglement of the steady state by computing the genuine multipartite negativity (GMN) (see \cref{appendix:Modeling:ent}) \cite{jungnitsch2011taming}. GMN is an entanglement witness based metric that is finite ($0 <\mathrm{GMN}\leq 0.5$) in the presence of genuine multipartite entanglement and 0 otherwise. We plot GMN against drive duration in the \cref{fig:ddre}d inset, observing a steady state value of $6.3^{+0.8}_{-0.5}\times10^{-2}$ (95$\%$ confidence interval). The ideal target of this protocol is a pure multipartite-entangled dark state, inaccessible to reciprocal waveguide schemes, which can prepare only pairwise (dimer) entanglement \cite{pichler2015quantum}. In practice, finite Purcell factors and parasitic decoherence leave the prepared state mixed; nonetheless, the witnessed genuine multipartite entanglement confirms that the protocol operates in this regime, with the residual mixedness consistent with these known imperfections (see \cref{appendix:Modeling:ddre fits}).

\section*{Photon bound states} 

\begin{figure*}[t!]
\centering
\includegraphics[width=1\linewidth]{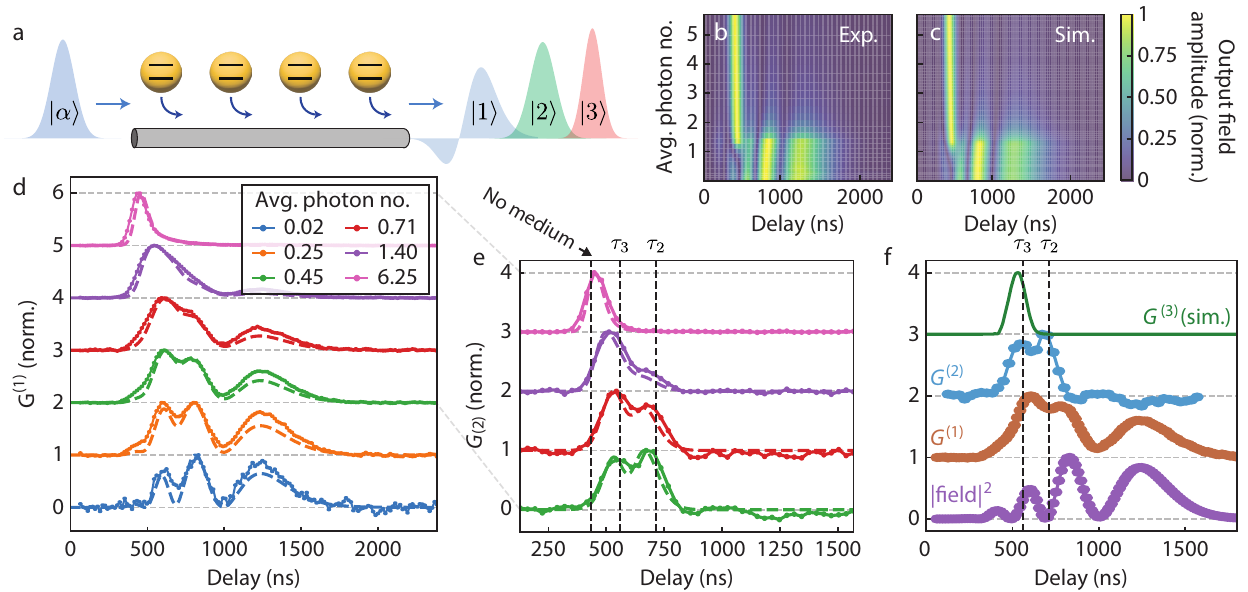}
\caption{\textbf{Photon bound states.} (a) Photon sorting schematic: number state components of a coherent pulse incident on the qubit chain are temporally separated at the output, with higher-order Fock states experiencing lower group delay. (b) Measured and (c) simulated output field amplitude, as a function of the input average photon number. Each trace is normalized against its own maximum for clarity. (d) Measured normalized $G^{(1)}(t,t)$. (e) Measured $G^{(2)}(t,t)$. $\tau_{j}$ represents the expected time of arrival for the $j$'th Fock-state (see \cref{appendix:Modeling:photon sorting}). (f) Comparison of different order correlators for average photon number of 0.45.} 
\label{fig:photon_sorting}
\end{figure*}

The driven-dissipative entanglement generation examined so far provide a probe of the cascaded network through the emitter degrees of freedom. The photons propagating through the waveguide provide a complementary probe. A chain of cascaded quantum emitters behaves as an extended non-linear medium for traveling photons, in which the single-photon non-linearity of each emitter accumulates coherently along the chain as a pulse propagates without the interference between counter-propagating photons that complicates reciprocal media. This distributed interaction gives rise to photon bound states, correlated scattering eigenstates in which a well-defined number of photons co-propagate as a unit \cite{mahmoodian2020dynamics, liang2018observation}. Such bound states are a manifestation of strong photon-photon interactions, a long-sought resource for photonic quantum logic \cite{shapiro2006single, gea2010impossibility, levy2025passive} and metrology \cite{paulisch2019quantum} that remains difficult to realize in any platform \cite{chang2014quantum}. We access this physics directly, scattering weak coherent pulses off the four-qubit chain and resolving the scattered light into bound states sorted in time by their photon number.

This temporal sorting originates in the photon-number dependence of the group delay. An $n$-photon Fock state scattered through a quantum emitter experiences a Wigner delay inversely proportional to the square of its photon number, $\sim 1/n^{2}$~\cite{mahmoodian2020dynamics}. The first factor of $1/n$ arises from the qubit absorbing a single photon from the pulse, and the second from the stimulated emission of that photon, an event that can be induced by any of the remaining non-absorbed photons and therefore carries a $\sim n\,\Gamma_\mathrm{1D}$ dependence. Because higher photon numbers are delayed less, a weak coherent pulse scattered off the chain is sorted in time into its constituent Fock states, as illustrated in \cref{fig:photon_sorting}a. 	While the photon-number dependence of the Wigner delay is present in scattering off a single emitter, the differential delay between Fock components always remains smaller than the minimum pulse width that the emitter can efficiently absorb \cite{tomm2023photon}. By cascading four qubits, we instead access a regime in which the bound states are sorted into distinct, well-resolved arrivals. 

We scatter coherent pulses of varying power off the four cascaded qubits and measure the output radiation (see \cref{appendix:Methods:field_tomography}, \cref{appendix:Methods:ampnoise}). The measured output field is shown in \cref{fig:photon_sorting}b. For mean photon numbers $n \ll 1$, the field is nearly independent of power, reflecting negligible qubit excitation. As the power increases, the qubits begin to saturate, producing a power-dependent output field that eventually approaches the classical soliton regime \cite{mahmoodian2020dynamics}. Simulations agree well with the measurement, as shown in \cref{fig:photon_sorting}c (see \cref{appendix:Modeling:photon sorting}).

We can identify individual bound states by measuring the same-time correlators, $G^{(j)}(t, t) \sim \langle [ \hat{a}^{\dagger}(t) ]^j [\hat{a}(t) ]^j\rangle$, where $\hat{a}(t)$ is the time-dependent output field. The correlator $G^{(j)}(t,t)$ represents the arrival of at least $j$ coincident photons to the detector at time $t$. The measured $G^{(1)}(t,t)$, or intensity, is shown in \cref{fig:photon_sorting}d for different input powers. We see that for increased input power (larger average $n$), the wavepacket experiences less group delay ($\sim1/n^2$) and arrives at the detector earlier. The second-order correlator, $G^{(2)}(t,t)$, is plotted in \cref{fig:photon_sorting}d, where we observe two distinct peaks. These peaks correspond to the two- and three-photon bound states, with the three-photon state arriving earlier and growing with increased input power. Observed group delays coincide with simulations, as denoted by $\tau_{2}$ and $\tau_{3}$ in \cref{fig:photon_sorting}e.

Comparing the squared field, $G^{(1)}(t,t)$, and $G^{(2)}(t,t)$ for a fixed input power in \cref{fig:photon_sorting}f provides several insights. First, the squared field differs greatly from the intensity $G^{(1)}(t,t)$, indicating that the output is no longer a coherent state. Next, the peak in $G^{(1)}(t,t)$ at $\sim 1200$ ns is absent in $G^{(2)}(t,t)$, revealing the single-photon component. Lastly, comparing $G^{(2)}(t,t)$ with the simulated $G^{(3)}(t,t)$ reveals the two-photon component, which is delayed with respect to the three-photon wavepacket. We do not directly measure $G^{(3)}(t,t)$, as amplifier noise makes it prohibitively costly (see \cref{appendix:Methods:field_tomography}). Together, these measurements demonstrate the use of a chiral chain as a medium for controlled, many-body quantum optics of light, with applications in metrology, computation, and simulation.

\section*{Conclusions and outlook}


This work establishes deterministic cascaded quantum systems as an experimental setting for many-body quantum optics. Within this setting, we observe dissipative many-body states of the emitters and strongly correlated transport of photons, complementary regimes predicted within the cascaded-systems framework introduced three decades ago~\cite{10.1103/physrevlett.70.2273, 10.1103/physrevlett.70.2269}. Looking ahead, two main directions follow, to which this system provides a concrete entry point.

The first direction is to utilize unidirectional qubit–bath interactions for quantum computing. A shared waveguide couples every emitter through traveling photons, naturally providing long-range all-to-all connectivity without the limitations of conventional superconducting interconnects \cite{niu2023low, kurpiers2018deterministic, Axline_2018, PhysRevLett.120.200501}. Combined with photon-mediated gate protocols \cite{mcintyre2025protocols}, this connectivity offers a natural route to modular superconducting processors with reduced error-correction overhead \cite{yoder2025tour}. Alternatively, the nonlinear emitter-photon interaction enables deterministic processing of the photons themselves, including photon addition and subtraction \cite{lund2024subtraction} and conditional gates between co-propagating photons \cite{levy2025passive, levy2025engineering}. Realizing these applications will require Purcell factors on the order of $10^4$ \cite{paulisch2016universal}, a requirement set primarily by thermal waveguide occupation, which can relaxed by modest reductions in temperature or increases in qubit frequency (see \cref{appendix:Device:chirality tuneup}). 

The second direction is the exploration of emergent many-body phenomena in large chiral arrays. The dimerized and multipartite steady states demonstrated here for four qubits are the smallest members of a family of pure steady states predicted for driven chiral chains \cite{stannigel2012driven, pichler2015quantum}. Extending these systems to tens of emitters opens genuinely many-body regimes, where symmetry-protected topological states can be stabilized \cite{pichler2015quantum, chu2025reconfigurable}. Large chiral arrays may also host novel entangled phases, where bond dimension grows with system size \cite{clark2020observation}, as well as many-body solitons \cite{mahmoodian2020dynamics} and superradiant bursts \cite{cardenaslopez2023manybody}. The key remaining challenge is therefore scaling the present architecture to larger arrays, through higher on-chip qubit density and parametric pump generation hardware that grows efficiently with system size. Together with the capabilities demonstrated here, these advances will bring the full many-body landscape of cascaded quantum systems within experimental reach.

\section*{Acknowledgments}

The authors thank H. Luo, S. Mahmoodian, A. Bozkurt, A. Clerk, and W. Pfaff for helpful discussions. We also thank K. Villegas and C.J. Wu of Quantum Machines and D. Goins of Windfreak Technologies for their technical support. This work was supported by the Office of Naval Research (award number:\, N00014-24-1-2052), the Air Force Office of Scientific Research (award number:\, FA9550-24-1-0354), the National Science Foundation (award number:\,1733907), and startup funds from Caltech's EAS division.  P.S.S. gratefully acknowledges support from the S2I-Gupta Fellowship. F.Y. gratefully acknowledges support from the NSF Graduate Research Fellowship. C.J. gratefully acknowledges support from the IQIM/AWS Postdoctoral Fellowship.

\bibliography{references}
\clearpage
\appendix
\section{Methods}
\label{appendix:Methods}

\subsection{Fabrication}
\label{appendix:Methods:fab}
Devices are fabricated on 1 cm $\times$ 1 cm high-resistivity (10 k$\Omega$-cm) silicon substrates. Bare chips are solvent cleaned in N-methyl-2-pyrrolidone at 150$^\circ$ C, acetone, and isopropyl alcohol. This is followed by oxygen plasma and buffer hydrofluoric acid (BHF) treatment. Electron-beam lithography is used to pattern structures in separate metal layers on the chip. Lithography steps are followed by electron-beam evaporation of metal, liftoff in N-methyl-2-pyrrolidone at 150$^\circ$ C for 1.5 hours, and surface treatments. Device layers are as follows. 
\begin{enumerate}[label=\roman*.]
    \item 150 nm thick niobium markers, deposited at 3 \text{\normalfont\AA}/s. After marker deposition, chips are treated with oxygen plasma, dipped in Pure Strip (stabilized H$_2$SO$_4$-H$_2$O$_2$ compound) at 60$^\circ$ C for 20 minutes, and treated with BHF. 
    \item 120 nm thick aluminum ground plane, control lines, waveguide, readout resonators, and qubit capacitors, deposited at 5 \text{\normalfont\AA}/s. Just prior to ground plane evaporation, chips are treated again with oxygen plasma and BHF. Another oxygen plasma treatment is performed after metal liftoff. 
    \item Josephson junctions evaporated (at 5 \text{\normalfont\AA}/s) using double angle evaporation and consisting of 60 nm and 120 nm layers of aluminum, with 15 minutes of static oxidation between layers. Just prior to junction evaporation, chips are treated with oxygen plasma and vapor hydrofluoric acid (VHF). Oxygen plasma treatment is repeated after metal liftoff. 
    \item 150 nm thick aluminum band-aids and air-bridges, deposited at 5 \text{\normalfont\AA}/s. Band-aids ensure electrical contact between Josephson junctions and qubit capacitors. Air bridges are used to ensure the suppression of the slot-line modes in the waveguide and control lines \cite{Chen2014Feb}. Air-bridges are patterned using grey-scale electron-beam lithography and developed in a mixture of isopropyl alcohol and de-ionized water, followed by 10 minutes of reflow at 105$^\circ$ C. Electron beam evaporation of the band-aid/bridge layer is preceded by 7 minutes of Ar ion milling. Oxygen plasma treatment is performed before evaporation and after liftoff. 
\end{enumerate}

\begin{figure}[htbp]
\centering
\includegraphics[width=1\linewidth]{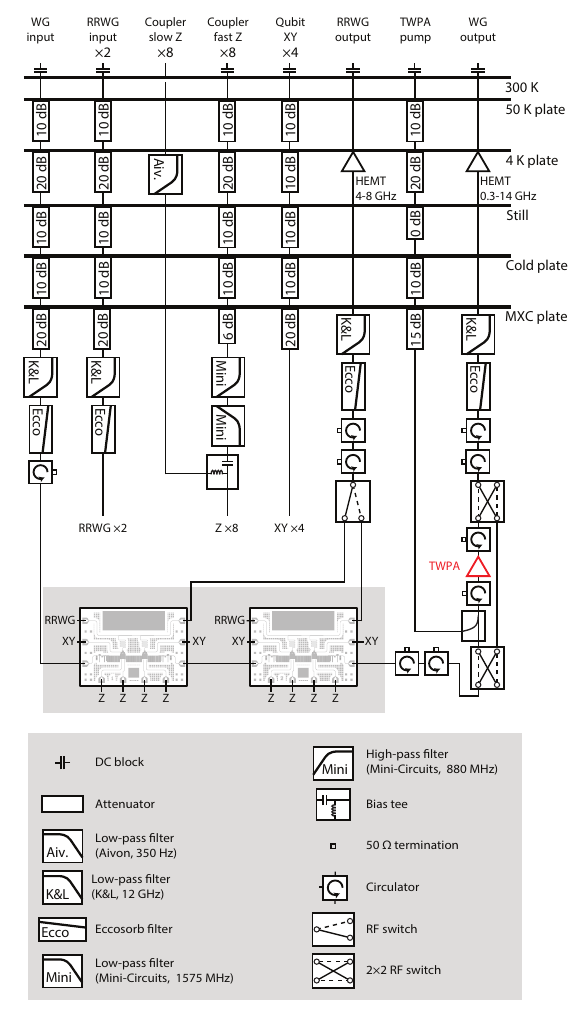}
\caption{\textbf{Schematic of dilution fridge wiring.}}
\label{fig:Fridge_wiring}
\end{figure}

\begin{figure}[htbp]
\centering
\includegraphics[width=1\linewidth]{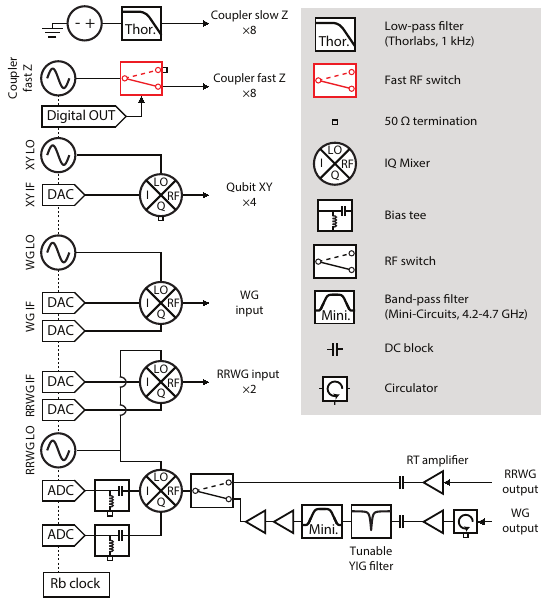}
\caption{\textbf{Schematic of room temperature setup.} }
\label{fig:RT_wiring}
\end{figure}

\subsection{Measurement setup}
\label{appendix:Methods:setup}

\subsubsection{Cryogenic setup}

Measurements are performed in a $^3$He/$^4$He dilution refrigerator. A schematic of the cryogenic measurement setup is shown in \cref{fig:Fridge_wiring}. Fabricated devices are wire-bonded to a PCB and placed in copper boxes. Boxes are mounted to the mixing plate (MXC) and cooled to a base temperature of 13 mK. The waveguide input line (WG input) is used to probe the coupled cavity array (CCA) waveguide and drive qubits coupled to the waveguide. It is attenuated at each temperature stage to minimize thermal noise. The total attenuation is 70 dB (2082-6418-$\square\square$-CRYO attenuators are used at 300 K, 4 K, still, and cold plate; a QMC-CRYOATT-20 attenuator is used at MXC). At the mixing plate, a 12 GHz K$\&$L low-pass filter (6L250-12000/T26000-OP/O) and Eccosorb filter reduce millimeter-wave and infrared noise, respectively. Next, a microwave circulator (LNF-CIC4$\_$8A) is inserted prior to the first chip input to minimize Fabry-Perot reflections. The two chips are connected via a 45 cm superconducting aluminum cable (Hermerc Systems). The output of the second chip contains a bypass consisting of two 2$\times$2 switches (Radiall R577423100LP) to allow for optional use of a traveling wave parametric amplifier (TWPA) provided by MIT Lincoln Laboratory. A total of 6 circulators are distributed in this output line to reduce reflections, and Eccosorb and K$\&$L filters are included after the TWPA and bypass. The output line is amplified by a high electron mobility transistor amplifier (HEMT, LNF-LNC0.3$\_$14B) at the 4 K stage. Two readout lines (RRWG input) are used for qubit state readout on the two chips. Similar to the waveguide input line (WG input), each line is attenuated by 70 dB and contains K$\&$L and Eccosorb filters. At the mixing plate, readout output lines are connected to an RF switch (Radiall R573423600), circulators, a K$\&$L filter, and an Eccosorb filter. Readout signals are amplified by a HEMT (LNF-LNC4$\_$8C) at 4 K. Coupler qubits embedded in the waveguide are frequency-tuned and frequency-modulated by slow Z and fast Z lines, respectively. Slow Z lines provide a constant DC bias to flux-tune coupler frequencies, and are low-pass filtered at the 4 K stage (Aivon Therma-uD25-G2F with 350 Hz cutoff frequency). The fast Z lines are attenuated by 56 dB and contain low-pass (Mini-circuits VLF-1575+) and high-pass filters (Mini-circuits VHF-880+) at the mixing plate. Slow and fast Z lines are then combined at AC-coupled bias tees (Mini-circuits ZFBT-4R2GW+, QMC-CRYOTEE-0.218). Emitter qubits are equipped with XY drive lines that are attenuated by a total of 60 dB. DC blocks (Centric RF CD9519) are included in all lines at room temperature.

\subsubsection{Room temperature setup}

The room temperature measurement setup is shown in \cref{fig:RT_wiring}. A low noise, multi-channel DC source (QDevil, QDAC) provides current biases for slow Z lines, and each DC channel is low-pass filtered (Thorlabs EF110, 1 kHz cutoff). RF sources (Windfreak SynthHD) are used to drive coupler fast Z lines. Fast Z lines are toggled using fast microwave switches (Mini-circuits ZYSWA-2-50DR+) controlled by the digital outputs of a Quantum Machines OPX+ controller. For elastic scattering measurements, the WG input line is driven by a vector network analyzer (VNA, Agilent N5242A). For subsequent experiments, the arbitrary waveform generator (AWG) of a Quantum Machines OPX+ is used to drive XY lines, the WG input line, and readout lines. Here, intermediate frequencies (IF) generated by the AWG are up-converted using IQ mixers (MMIQ-0218LXPC) and local oscillator (LO) tones provided by separate RF sources (Windfreak SynthHD, BNC 855). The RRWG and WG outputs are amplified at room temperature (RRWG: Wantcom WBA2080-35A, WG: Narda-MITEQ LNA-30-04000800-07-10P); amplifiers are DC isolated with DC blocks. The WG output also includes a circulator to suppress back-reflections, a tunable YIG filter (MLBFR-0212 band-reject) to remove the TWPA pump, and two additional amplifiers (Mini-circuits ZX60-02203+) preceded by a band-pass filter (Mini-circuits VBF-4440+, 4.2-4.7 GHz) to prevent amplifier saturation by noise. Depending on the measurement, an RF switch connects either the RRWG output or WG output to the RF port of an IQ mixer, which is down-converted and measured at the analog-to-digital converters of the OPX+. Bias tees (Mini-circuits ZFBT-6G+) are used to suppress high frequency noise prior to the ADC input. 

Not shown in \cref{fig:RT_wiring} is the TWPA pump tone and emitter qubit AC-Stark shift tones (added to XY lines when needed), which are both driven by Rohde and Schwarz SMB100A sources. Inelastic scattering measurements (used for calibration of amplifier noise, see \cref{appendix:Methods:ampnoise}) are performed by driving a single qubit via the WG input and measuring the emission with a spectrum analyzer (Rohde and Schwarz FSV3013).

\subsection{State tomography via dispersive readout}
\label{appendix:Methods:state tomography}

We perform quantum state tomography via dispersive readout to recover the full density matrices of stabilized four-qubit states. The density matrix for an $N$-qubit state contains $2^{2N}$ elements. Therefore, to uniquely reconstruct the density matrix, we must perform $4^N = 2^{2N}$ measurements of multi-qubit Pauli operators. For $N$ qubits, the multi-qubit Pauli operators are defined as all the possible $N$ element tensor products of the single qubit Pauli operators $X=\hat{\sigma}^x$, $Y=\hat{\sigma}^y$, $Z=\hat{\sigma}^z$, and the identity operator $I$. For $N=4$ qubits, this includes all single-qubit (i.e. $\braket{IXII}$), two-qubit (i.e. $\braket{IXYI}$), three-qubit (i.e. $\braket{YXIZ}$), and four-qubit (i.e. $\braket{XZXX}$) Pauli operators, where operators are listed in order by qubit position in the chain. There are a total of $4^4 =256$ such operators for a four-qubit state. To measure the Pauli operator $X$ $(Y)$ for a single qubit, we perform a $\pi/2$ rotation along the y (x) axis of the Bloch sphere, followed by dispersive readout. To measure $Z$, no rotation is needed prior to dispersive readout. State tomography then involves preparing the state of interest multiple times, and for each prepared state, performing the necessary qubit rotations for a desired multi-qubit Pauli operator, followed by dispersive readout of each qubit. We comment that, in our experiment, the qubits are de-coupled from the waveguide during readout, preventing loss of the quantum state via waveguide decay. We turn off qubit-waveguide coupling using fast microwave switches to shut off parametric coupler drives. In the following section, we detail the tune-up and calibration of the state preparation and tomography. 

\subsubsection{Readout}

Each qubit is coupled dispersively to a readout resonator, and the frequency of the resonator shifts (dispersive shift of $\pm\chi$) depending on if the qubit is in the ground or excited state. This allows us to measure the qubit state by probing the readout resonator with a microwave pulse. We perform frequency multiplexed qubit readout, sending pulses of a different frequency to each readout resonator simultaneously (see \cref{tab:table1} for parameters). Readout pulses are synthesized by an AWG (see \cref{appendix:Methods:setup}, \cref{fig:RT_wiring}). We optimize the signal-to-noise ratio (SNR) of the readout by sweeping the frequency, power, and duration ($\tau_\mathrm{RO}$) of readout pulses. We find optimal performance at $\tau_\mathrm{RO} \approx 1$ $\mu$s, corresponding to a readout SNR of $\approx0.6$ for each qubit (see \cref{tab:table1}).

We observe drifts in the phase and amplitude of the readout signal on an $\sim$hour time-scale, which we attribute to slow resonator dephasing. This effect can compromise the state tomography by introducing mixedness into the reconstructed density matrix. To correct for this, we periodically calibrate the readout signal by measuring and fitting Rabi oscillations of each individual qubit. To measure Rabi oscillations, each qubit is excited by pulses of varying duration via its XY line. For each pulse duration, we perform dispersive readout. The resulting Rabi oscillations are fit to theory to calibrate the expected signal for qubit ground and excited states.

\subsubsection{Local rotations}

Dedicated XY lines are used to perform individual qubit rotations. To tune up local rotations, we calibrate qubit frequencies, pulse amplitudes, and pulse durations. We obtain qubit frequencies via pulsed spectroscopy. Pulse amplitudes and durations are calibrated by measuring and fitting qubit Rabi oscillations. We then fine-tune settings by applying different combinations of two consecutive single qubit gates ($X$, $Y$, $X/2$, $Y/2$), comparing measured and expected results following established techniques \cite{reed2013entanglement}. All $\pi/2$ $(\pi)$ pulses are kept at $\tau_\mathrm{XY} = $ 64 (128) ns.

We observe classical crosstalk between XY lines and qubits on the same chip; A tone applied to the first qubit's XY line parasitically drives the second qubit. To compensate for this effect, we calibrate correction tones to destructively interfere with parasitic drives. Therefore, to locally drive a single qubit, we apply two microwave tones: the first tone is applied to the target qubit's XY line, and the second tone is applied to the adjacent qubit's XY line to cancel crosstalk. We do not observe any classical crosstalk between qubits on different chips.

\subsubsection{State reconstruction}

For a four-qubit density matrix, we measure $4^4=256$ multi-qubit Pauli operators $\hat{P_j}$. To reconstruct the density matrix from measurements, we perform maximum likelihood estimation (MLE) following standard procedures \cite{eichler_thesis, vinicius_thesis}. Assuming a quantum state $\rho$, the measurement outcomes for a given operator $\hat{P}_j$ should follow a probability distribution with mean $\mathrm{Tr}(\hat{P}_j \rho)$ and variance $v_j$. Provided $M$ measurements (and large enough $M$), the sampling distribution will be Gaussian by the central limit theorem, with the same mean and a reduced variance, $v_j/M$. For a state $\rho$, the probability of measuring a sample mean $\langle \bar{P}_j\rangle$ is $p(\langle \bar{P}_j\rangle|\rho)$, which takes on the form given below. 

\begin{equation}
p(\langle \bar{P}_j\rangle|\rho) \propto e^{-|\langle\hat{P}\rangle-\mathrm{Tr}(\hat{P}_j\rho)|^2 / (v_j/M)}
\label{MLE}
\end{equation}

\noindent For large $M$, we can replace $v_j/M$ by the sample variance, $\bar{v}_j$. We then define the negative log-likelihood functional below, accounting for all measurements of multi-qubit Pauli operators. 

\begin{equation}
-\log \mathcal{L} (D|\rho) = \sum_{j=1}^{4^N}|\langle \bar{P}_j \rangle - \mathrm{Tr}(\hat{P}_j\rho)|^2 / (\bar{v}_j)
\label{logMLE}
\end{equation}

\noindent Here, $D$ is the set of sample means and variances $\{(\langle \bar{P}_j \rangle, \bar{v}_j)\}_{j=1}^{4^N}$. Minimizing the functional over possible density matrices produces the state $\rho$ with highest likelihood to result in the measured observables. Combining the functional with the condition that $\mathrm{Tr}(\rho)=1$ and $\rho$ be positive semi-definite ($\rho>0$) fully defines the optimization problem for recovering the quantum state.

To obtain bounds on metrics such as pair-wise concurrences or fidelities, we re-sample the Pauli operators from normal distributions derived from the measured means and variances. For each re-sampled dataset, we perform MLE to reconstruct a new density matrix. We then calculate the metric of interest for each new density matrix to build a sampling distribution for the target metric, from which we determine standard deviations and confidence intervals. 

\subsubsection{Calibration and experiment}

\begin{figure}[htbp]
\centering
\includegraphics[width=0.9\linewidth]{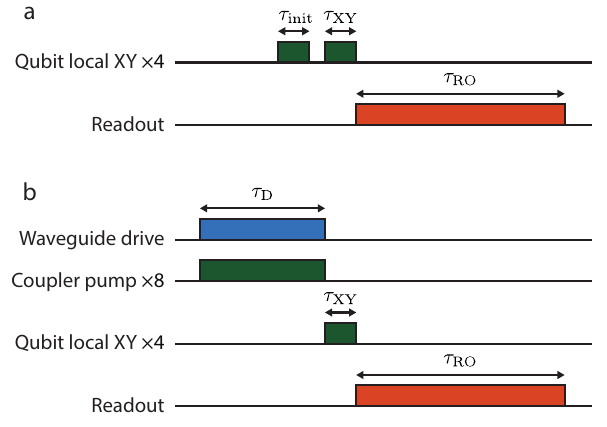}
\caption{\textbf{State tomography pulse sequences.} Initialization and measurement of (a) cardinal states and (b) stabilized entangled states. In both cases, local drives are used rotate the state according to a target Pauli string, followed by dispersive readout. In (a), states are prepared by local drives. In (b), states are prepared by a waveguide drive, with qubit-waveguide couplings turned on. After time $\tau_\mathrm{D}$, the waveguide drive and qubit-waveguide couplings are turned off, and local rotations and readout are performed.}
\label{fig:pulses}
\end{figure}

To recover a density matrix from experiments, we initialize the state, perform local qubit rotations depending on a target Pauli operator, and then perform dispersive readout (see \cref{fig:pulses}a,b). We repeat this procedure $M \times 3^N$ times, corresponding to measuring each four-qubit Pauli operator exactly $M$ times. Because single-, two-, and three-qubit Pauli operators are sub-sets of the four-qubit Pauli operators, we can obtain these measurements simultaneously (without additional measurements). For example, initializing and rotating a target state to measure the $XXXX$ operator also allows us to measure sub-operators such as $XIII$, $IXXI$, and $XXIX$. This allows us to obtain $4^N-1$ Pauli operators, where the identity operator $IIII$ is omitted.  

To benchmark the state tomography, we first prepare and measure six separable joint qubit states: $|$$\pm x\rangle^N$, $|$$\pm y\rangle^N$, and $|$$\pm z\rangle^N$, similar to procedures in previous works \cite{irfan2025autonomous}. We call these the cardinal states, and for four qubits they are defined as $|$$\pm a\rangle^4 = |$$\pm a\rangle \otimes |$$\pm a\rangle \otimes |$$\pm a\rangle \otimes |$$\pm a\rangle$ for $a=x,y,z$. We prepare cardinal states using local rotations: $|$$-z\rangle$ is the ground state, $|$$+z\rangle$ is prepared by applying $\pi$ pulses to all qubits, and $|$$\pm x\rangle$ and $|$$\pm y\rangle$ are prepared by applying $\pm \pi/2$ pulses to all qubits along the $y$ and $x$ axes, respectively. The full pulse sequence for this calibration is shown in \cref{fig:pulses}a. Taking $M=5\times10^5$ averages, we obtain an average fidelity of $96.0^{+2.8}_{-4.4}$$\%$ ($95\%$ confidence interval) over all cardinal states. 

After calibrating the tomography procedure, we proceed to the stabilization experiment. Here, we drive the waveguide-coupled chiral qubits by applying a microwave tone to the waveguide. At some variable time $\tau_\mathrm{D}$, we turn off both the waveguide drive and the qubit-waveguide couplings to read out the quantum state. This sequence is shown in \cref{fig:pulses}b. We take $M=5\times10^5$ averages for this experiment. 

For the dimerization (multi-partite) experiment, we use a 4.646 (4.644) GHz stabilization tone at an on-chip power of -141.0 (-145.5) dBm. This yields an estimated Rabi drive of 1.8 (1.1) MHz in the dimer (multi-partite) case (using $\Gamma_\mathrm{1D}^f=2$ MHz). In both experiments, we vary the drive power to maximize a relevant figure of merit. For dimers, we maximize the pair-wise concurrences between qubits 1 and 2 and qubits 3 and 4. For the multi-partite entangled state, we maximize the genuine multipartite negativity (GMN, see \cref{appendix:Modeling:ent}).

We comment that, for high waveguide drive powers, coupler transmons can saturate, leading to reduction in waveguide transmission and potentially affecting emitter dynamics. However, we do not observe any waveguide saturation below drive powers of -125.0 dBm. Because our stabilization experiments are performed at much lower drive powers, we should be un-affected by these non-idealities. This is confirmed by the agreement between measured states and simulations (see \cref{fig:ddre}, \cref{appendix:Modeling:ddre fits}).

The qubit frequency conversion that accompanies our waveguide coupling scheme necessitates careful accounting of accumulated phase during the experiment. For a qubit with a natural frequency of $\omega_\mathrm{Q}$, the emitter interacts with the waveguide at frequency $\omega_\mathrm{Q}+\Delta$, where $\Delta$ is the frequency of the input coupler pump (see \cref{appendix:Design:CMT}). Therefore, there is a frequency difference of roughly $\Delta \approx 1 $ GHz between the waveguide drive ($\approx\omega_\mathrm{Q}+\Delta$) and local rotation pulses ($\omega_\mathrm{Q}$). (See \cref{tab:table1} for details). This large frequency difference can result in scrambling of the frame in which local rotations are applied, relative to the waveguide drive frame. In our experiment, we remove these harmful effects by carefully monitoring accumulated phases and applying corrective virtual Z gates to individual qubits.

\subsection{Field tomography}
\label{appendix:Methods:field_tomography}

To observe photon bound states, we excite the four-qubit chain with Gaussian pulses (voltage amplitude $\propto e^{-(\frac{t}{2\sigma})^2}$) via the waveguide input and measure same time correlators of the output field, $G^{(j)}(t,t)$. We use pulses of varying power and constant width $\sigma = 93$ ns, giving $1/2\pi\sigma = 1.7$ MHz, which is commensurate with individual qubit linewidths. This choice of pulse width roughly maximizes the scattering probability into two and three-photon bound states \cite{mahmoodian2020dynamics}. The output field correlators are measured using field tomographic techniques, which are described in detail in our previous work \cite{shah2024stabilizing}. 

Scattered microwave radiation must be amplified prior to detection, resulting in injection of noise photons. The effective noise photon number for our measurement chain is $n_\mathrm{TWPA} = 4.35$ (see \cref{appendix:Methods:ampnoise}). This exponentially increases the experimental cost of field measurements for higher-order correlators; the number of measurements required to obtain a correlator of order $j$ with a given SNR scales as $(1+n_\mathrm{TWPA})^{2j}$ \cite{eichler_thesis}. For measurements of $G^{(2)}(t,t)$, we vary the number of averages with the input power from $6.5\times10^{6}$ to $2.5\times10^{10}$ shots (from highest to lowest powers), with the longest measurement taking 53 hours. In particular, the $G^{(2)}(t,t)$ for an average photon number of 0.45 (presented in \cref{fig:photon_sorting}f) was obtained using $2.5\times10^{10}$ shots taken over 53 hours. Measuring the $G^{(3)}(t,t)$ at this input power setting (with the same SNR) would require $\sim$$29$ times more averaging, with an expected measurement time of 1500 hours, which we deemed infeasible. 


\subsection{Amplifier noise}
\label{appendix:Methods:ampnoise}

For transmission and photon sorting measurements, the output signal from the CCA waveguide is amplified at the mixing stage using a traveling wave parametric amplifier (TWPA), as shown in \cref{fig:Fridge_wiring}. To measure the added noise of the TWPA, we refer the amplifier noise to the emission of a single qubit into the waveguide using a resonance fluorescence measurement. Here, we turn on the waveguide coupling for the furthest downstream qubit (qubit 4) and drive the qubit on resonance with a single tone through the waveguide. We use qubit 4 because it is closest to the TWPA in the measurement chain. Therefore, the extracted noise photon number will not account for loss in the waveguide prior to (upstream of) qubit 4. Resonance fluorescence is measured using a spectrum analyzer (see \cref{appendix:Methods:setup}). For a single qubit with bidirectional coupling to the waveguide, we expect to measure peak emission of 0.5 photons. When referenced to the amplifier noise, this yields an effective noise photon number $n_\mathrm{TWPA} = 4.35$, corresponding to a temperature of $T_\mathrm{TWPA} = 936$ mK. We note that $n_\mathrm{TWPA}$ could be lowered by removing the 2$\times$2 switch prior to the TWPA (see \cref{fig:Fridge_wiring}), reducing the waveguide loss between qubit 4 and amplifier.

Signal from the readout waveguide is amplified using a HEMT at the 4 K stage. The effective HEMT noise photon number is extracted in a qubit resonance fluorescence measurement from a prior cooldown \cite{shah2024stabilizing}. This yields $n_\mathrm{HEMT} = 13$ photons, corresponding to a noise temperature of 4.1 K.

\section{Device overview}
\label{appendix:Device}

\subsection{Device summary}
\label{appendix:Device:summary}

\subsubsection{Waveguide and couplers}

\begin{figure}[htbp]
\centering
\includegraphics[width=1\linewidth]{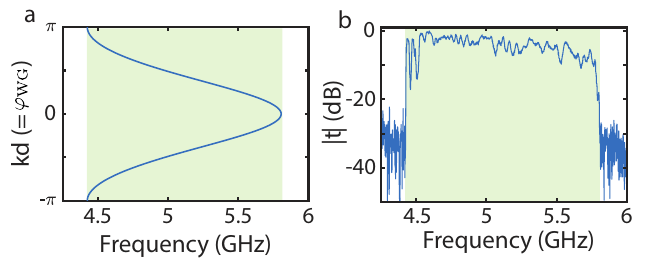}
\caption{\textbf{Coupled cavity array passband.} (a) Dispersion relation of the coupled cavity array (CCA) waveguide (from theory) (b) Measured CCA transmission, showing a passband from $4.4-5.8$ GHz and bandgaps outside of this region. }
\label{fig:passband}
\end{figure}

The coupled cavity array (CCA) waveguide consists of a chain of resonant, capacitively coupled, lumped-element superconducting resonators (\cref{fig:Fig1}c, blue). Each resonator is coupled to its nearest neighbors with interaction strength $J$. The CCA waveguide can be described by a tight-binding Hamiltonian, with the dispersion relation given by $\omega(k) = \omega_0 + 2J\cos(kd)$, as plotted in \cref{fig:passband}a ($\omega_0$ is the bare cavity frequency, $k$ is the wavevector, $d$ is the lattice constant). It features a transmissive passband of width $4J$ (highlighted in green) and a strong suppression of the photonic density of states outside the passband. \cref{fig:passband}b shows the measured CCA transmission, with a passband of $4.4-5.8$ GHz. For a detailed discussion of the CCA design, see \cref{appendix:Design:circuit}. 

Two adjacent flux-tunable coupler transmons are capacitively coupled to each emitter qubit and embedded in the CCA waveguide (\cref{fig:Fig1}c, purple). The couplers are impedance matched to the CCA resonators and contain SQUID loops with asymmetric Josephson junctions. To tune up the device, the coupler transmons are DC flux biased (using Z-lines pictured in \cref{fig:Fig1}d, green) to be approximately resonant with the resonators in the CCA waveguide. This is done in an iterative process to maximize the passband transmission as we sweep the frequency of each individual coupler. Coupler parameters are listed in \cref{tab:couplers}. 

\begin{table*}[htpb]
\begin{ruledtabular}
\begin{tabular}{lcccc}
\textrm{Parameter}&
\textrm{Max. frequency (GHz)}&
\textrm{Min. frequency (GHz)}&
\textrm{Junction asymmetry}\\
\colrule
Coupler 1, left & 6.210 & 3.462 & 1.94\\
Coupler 1, right & 6.344 & 3.232 & 1.73\\
Coupler 2, left & 6.239 & 3.221 & 1.76\\
Coupler 2, right & 6.299 & 3.608 & 2.02\\
Coupler 3, left & 6.106 & 3.087 & 1.72\\
Coupler 3, right & 6.133 & 3.421 & 1.94\\
Coupler 4, left & 6.068 & 3.170 & 1.79\\
Coupler 4, right & 5.931 & 3.399 & 2.02\\
\colrule
\end{tabular}
\end{ruledtabular}
\caption{ Parameters for coupler transmons. Couplers anharmonicity is -115 MHz (simulated).}
\label{tab:couplers}
\end{table*}

\subsubsection{Emitter qubits}

The emitter qubits are identical fixed-frequency transmons containing a single Josephson junction, with natural frequencies outside of the transmissive passband of the waveguide ($\omega_\mathrm{Q}\approx3.2-3.4$ GHz). We use fixed-frequency qubits to mitigate the effect of flux noise. To enable parametric coupling between the emitter qubits and the waveguide, the coupler elements are frequency modulated by applying microwave tones to their respective Z-lines. For a drive tone of frequency $\Delta$, the emitter qubit then breaks into frequency components - ``sidebands'' - at $\omega_\mathrm{Q} + n\Delta$, where $n\in \mathbb{Z}$. In our design, the first blue sideband ($n=1$, $ \omega_\mathrm{Q1} = \omega_\mathrm{Q} + \Delta$) constitutes the chiral qubit and lies in the waveguide passband. All other sidebands are suppressed by the CCA bandgap. Because the chiral qubit frequency can be chosen using $\Delta,$ the use of fixed-frequency emitter qubits does not compromise the frequency tunability of the waveguide-coupled chiral qubits.

Each emitter is equipped with a control (XY) line used to excite the qubit and a readout resonator for dispersive measurement. We measure qubit lifetimes ($T_1$), Ramsey coherence times ($T_2^*$), spin-echo coherence times ($T_{2E}$) with dispersive readout. For all time domain measurements, coupler frequencies are biased to the center of the waveguide passband. The qubit anharmonicities ($\alpha$) are measured using two-tone spectroscopy; the readout resonator transmission is monitored continuously as a pump tone applied to the qubit XY line and swept in frequency. The readout resonator dispersively shifts when the pump tone coincides with $\omega_\mathrm{Q}$ and $\omega_\mathrm{Q} - \alpha/2$, corresponding to resonantly driving the qubit $g\rightarrow e$ transition and two-photon transition ($g\rightarrow f$), respectively. $\alpha$ is the qubit anharmonicity; and $g$, $e$, and $f$ are the ground, first excited, and second excited states of the qubit. Measured qubit parameters are provided in \cref{tab:table1}. 

\subsubsection{Readout resonators}

Readout resonators are co-planar waveguide $\lambda/4$ resonators with capacitive claws designed for coupling to emitter qubits. They are inductively coupled to a readout waveguide separate from the CCA waveguide, and detuned in frequency to allow for frequency multiplexed dispersive readout. Readout resonator parameters are provided in \cref{tab:table1}. 

\begin{table*}[htpb]
\begin{ruledtabular}
\begin{tabular}{lcccc}
\textrm{Parameter}&
\textrm{Emitter Qubit 1}&
\textrm{Emitter Qubit 2}&
\textrm{Emitter Qubit 3}&
\textrm{Emitter Qubit 4}\\
\colrule
Frequency $\omega_\mathrm{Q}$ (GHz) & 3.425 & 3.423 & 3.283 & 3.256 \\
Lifetime $T_{1}$ ($\mu$s) & 12.1 & 12.1 & 14.6 & 15.8 \\
Ramsey $T^*_{2}$ ($\mu$s) & 4.7 & 7.2 & 9.0 & 12.1 \\
Spin echo $T_{2E}$ ($\mu$s) & 13.0 & 12.5 & 12.9 & 14.8 \\
Anharmonicity $\alpha$ (MHz) & -172.0 & -196.8 & -190.0 & -172.6 \\
\colrule
\textbf{Readout} \\
\colrule
Readout resonator frequency (GHz) & 4.367 & 4.349 & 4.329 & 4.312 \\
Readout resonator external decay rate (MHz) & 5.01 & 4.14 &  5.41 & 3.59 \\
Readout resonator internal decay rate (kHz) & 23.3 & 26.7 & 32.7 & 118.1 \\
Qubit-resonator detuning (MHz) & 942 & 926 & 1046 & 1056 \\
Dispersive shift $\chi$ (kHz) & 223 & 241 & 347 & 284 \\
Qubit-resonator coupling (MHz) & 30.7 & 28.8 & 40.4 & 39.2 \\
Readout pulse duration $\tau_\mathrm{RO}$ (ns) & 1100 & 1100 & 1300 & 1100 \\
Readout SNR & 0.67 & 0.67 & 0.60 & 0.66 \\
\colrule
\textbf{Elastic scattering, photon sorting} \\
\colrule
Modulation frequency (GHz) & 1.235 & 1.235 & 1.385 & 1.409 \\
AC Stark shift $\Delta_\mathrm{AC}$ (MHz) & 0 & 0 & 0 & -4 \\
Blue sideband frequency (GHz) & 4.657 & 4.657 & 4.657 & 4.661 \\
Qubit forward waveguide decay $\Gamma^{f}_\textrm{1D}$ (MHz) & 1.96 & 1.78 & 1.77 & 1.71 \\
Qubit parasitic decoherence $\Gamma^{b}_\textrm{1D} + \Gamma'$ (kHz) & 94.7 & 121.0 & 194.8 & 177.2 \\
Purcell factor $P_\mathrm{1D}$ & 20.8 & 14.7 & 9.1 & 9.7 \\
$\beta$-factor & 0.95 & 0.94 & 0.90 & 0.91 \\

\colrule
\textbf{Dimers} \\
\colrule
Modulation frequency (GHz) & 1.225 & 1.225 & 1.372 & 1.393 \\
AC Stark shift $\Delta_\mathrm{AC}$ (MHz) & 0 & 0 & 0 & 0 \\
Blue sideband frequency (GHz) & 4.646 & 4.646 & 4.648 & 4.644 \\
Qubit forward waveguide decay $\Gamma^{f}_\textrm{1D}$ (MHz) & 3.08 & 3.56 & 2.35 & 2.08 \\
Qubit parasitic decoherence $\Gamma^{b}_\textrm{1D} + \Gamma'$ (kHz) & 257.5 & 253.0 & 362.8 & 148.0 \\
Purcell factor $P_\mathrm{1D}$ & 12.0 & 14.1 & 6.5 & 14.1 \\
$\beta$-factor & 0.92 & 0.93 & 0.87 & 0.93 \\

\colrule
\textbf{Tetramer} \\
\colrule
Modulation frequency (GHz) & 1.225 & 1.225 & 1.366 & 1.393 \\
AC Stark shift $\Delta_\mathrm{AC}$ (MHz) & -2 & 0 & 0 & 0 \\
Blue sideband frequency (GHz) & 4.646 & 4.646 & 4.642 & 4.644 \\
Qubit forward waveguide decay $\Gamma^{f}_\textrm{1D}$ (MHz) & 2.24 & 2.94 & 1.86 & 1.79 \\
Qubit parasitic decoherence $\Gamma^{b}_\textrm{1D} + \Gamma'$ (kHz) & 202.1 & 230.0 & 176.4 & 200.5 \\
Purcell factor $P_\mathrm{1D}$ & 11.1 & 12.8 & 10.6 & 8.9 \\
$\beta$-factor & 0.92 & 0.93 & 0.91 & 0.90 \\

\end{tabular}
\end{ruledtabular}
\caption{\label{tab:table1}
Parameters for the emitter qubits and readout.}
\end{table*}

\subsection{Tuning up chiral qubits}
\label{appendix:Device:chirality tuneup}
\subsubsection{Chirality}

\begin{figure}[htbp]
\centering
\includegraphics[width=1\linewidth]{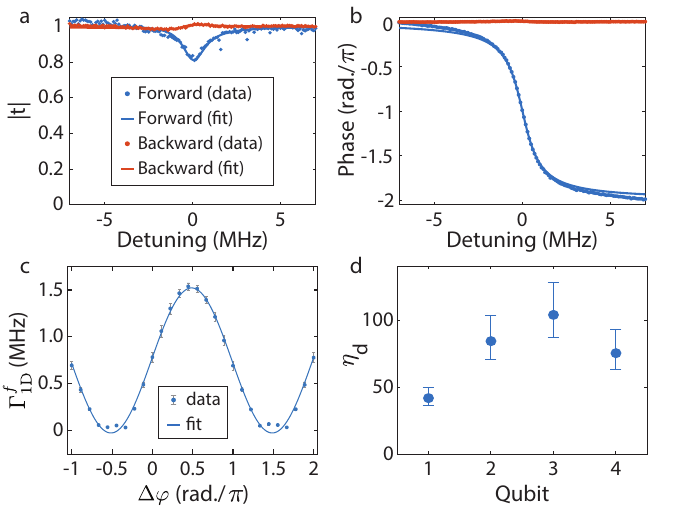}
\caption{\textbf{Qubit chirality.} (a) Transmission and (b) phase of a forward and backward chiral qubit, corresponding to two different settings of $\Delta\varphi$. The blue trace shows transmission for a forward chiral qubit, showing a transmission dip and $2\pi$ phase shift across the resonance. The red trace shows a backward chiral qubit, where the qubit decouples from the waveguide and becomes invisible to the probe tone. (c) Qubit forward decay rate $\Gamma^f_\mathrm{1D}$ extracted from fits to \cref{eq:transmission}, plotted against relative pump phase $\Delta\varphi$. $\Gamma^f_\mathrm{1D}$ has a sinusoidal dependence on $\Delta\varphi$. (d) Directionality ratio $\eta_\mathrm{d}$ for the four qubits in the cascaded chain, plotted with 95$\%$ confidence intervals.}
\label{fig:chirality}
\end{figure}

The natural frequencies of emitter qubits lie outside of the CCA passband ($\omega_\mathrm{Q}\approx3.2-3.4$ GHz), preventing radiative decay into the waveguide in the absence of parametric driving. Parametric pumps are applied to coupler Z-lines to modulate coupler frequencies, resulting in frequency conversion of the emitter into multiple ``sidebands'' at frequencies $\omega_\mathrm{Q} + n\Delta$, $n\in\mathbb{Z}$. We use $\Delta \approx1.2-1.4$ GHz, which places the first blue sideband frequency ($\omega_\mathrm{Q} + \Delta$) in the waveguide passband, turning on the qubit-waveguide coupling. Other sidebands lie outside of the passband and do not radiate into the waveguide. The choice of modulation frequency $\Delta$ determines the chiral qubit frequency.

Two neighboring coupler elements are coupled to each emitter (labeled left (L) and right (R)), and the relative phase between the two coupler drives $\Delta\varphi = \varphi_\mathrm{R} - \varphi_\mathrm{L}$ sets the chirality of the emitter qubit. We can tune the qubit chirality in situ to be forward or backward chiral, as shown in \cref{fig:chirality}a,b. The transmission coefficient for a single chiral qubit is given by

\begin{equation}
t = 1 - \frac{\Gamma^{f}_{\mathrm{1D}}}{i\delta + i\Gamma_{\mathrm{tot}}/2}
\label{eq:transmission}
\end{equation}

\noindent where $\Gamma_{\mathrm{tot}} = \Gamma^f_{\mathrm{1D}} + \Gamma^b_{\mathrm{1D}} + \Gamma'$ is the total decoherence of the qubit and $\delta = \omega_\mathrm{D} - \omega_\mathrm{Q}$ is the drive-qubit detuning \cite{joshi2023resonance}. The blue trace of \cref{fig:chirality}a,b shows the transmission and phase of a forward chiral qubit, where we observe a $2\pi$ phase shift across the resonance (and $\pi$ phase on resonance) as expected from \cref{eq:transmission}. The red trace shows the transmission of a backward chiral qubit, where the qubit decouples from forward propagating waveguide modes. This results in unit transmission and no phase response (Let $\Gamma_\mathrm{1D}^f=0$, resulting in $t\rightarrow1$). \cref{eq:transmission} is used to fit measured data in \cref{fig:chirality}a,b and \cref{fig:Fig1}e. We also use fits to \cref{eq:transmission} to obtain emitter Purcell factors $P_\mathrm{1D} = \frac{\Gamma_\mathrm{1D}^f}{\Gamma_\mathrm{1D}^b + \Gamma'}$, reported in \cref{tab:table1}. The transmission coefficient for a bidirectional qubit can be obtained by setting $\Gamma_\mathrm{1D}^f = \Gamma_\mathrm{1D}^b = \Gamma_\mathrm{1D}/2$, resulting in 

\begin{align}
t = 1 - \frac{\Gamma_{\mathrm{1D}}/2}{i\delta + i\Gamma_{\mathrm{tot}}/2}.
\end{align}

In contrast to the chiral qubit, a bidirectional qubit can never exhibit a $2\pi$ phase shift across its resonance, distinguishing the two cases. This is plotted in \cref{fig:Fig1}e in red. For completion, we also report $\beta = \Gamma_\mathrm{1D}^f/\Gamma_\mathrm{tot}= P_\mathrm{1D}/(P_\mathrm{1D}+1)$, which is an equivalent metric to Purcell factor.

As discussed in previous work \cite{joshi2023resonance}, the forward and backward waveguide decay rates are $\Gamma^{f/b}_\mathrm{1D} \propto 1 + \cos(\Delta\varphi \mp \varphi_{\mathrm{WG}})$, where $\varphi_\mathrm{WG}$ refers to the phase accumulated during waveguide propagation between the two couplers. For our device, $\varphi_\mathrm{WG} \approx \pi/2$. We observe the sinusoidal dependence of $\Gamma^f_\mathrm{1D}$ by sweeping $\Delta\varphi$ while probing the transmission in the forward direction, as shown in \cref{fig:chirality}c. The backward direction is not measured directly due to constraints in the measurement wiring (see \cref{fig:Fridge_wiring}). 

We extract the directionality ratio $\eta_\mathrm{d} = \Gamma_\mathrm{1D}^f/\Gamma^b_\mathrm{1D}$ of each qubit by fitting the transmission of each qubit in the forward and backward chiral phase setting to obtain $\Gamma^{f,b}_\mathrm{1D}$. The obtained $\eta_\mathrm{d}$ is plotted in \cref{fig:chirality}d and ranges from $\approx40-100$. The main limitation on the chirality extraction is likely impedance mismatch between the coupled cavity array and the 50 $\Omega$ waveguides used to probe the device; this leads to ``ripples'' in the transmission that can become confounded with a vanishingly small $\Gamma^b_\mathrm{1D}$, as discussed in previous works \cite{joshi2023resonance}. We expect that with modest changes to CCA settings, we can consistently obtain directionalities $\eta_\mathrm{d}$ exceeding 100, as shown in prior work \cite{joshi2023resonance}. 

\subsubsection{AC-Stark shift}

The frequency of a chiral qubit (shown in yellow in \cref{fig:Fig1}d) is set by the frequency of parametric drives applied to the couplers (shown in purple in \cref{fig:Fig1}d). The relative phase between the two pumps sets the qubit chirality. Tuning the frequency of a chiral qubit then involves adjusting the drive frequency, followed by re-calibrating the relative pump phases. To more conveniently adjust the frequency of a chiral qubit, we also use microwave tones applied to the qubit XY line (top left of \cref{fig:Fig1}d) to induce AC Stark shifts. \cite{carroll2022dynamics}. Qubit frequencies are Stark shifted according to

\begin{align}
\Delta_{\mathrm{AC}} = \frac{\alpha\Omega^2}{2\delta(\delta+\alpha)}
\label{eq:qubit-basis-hamiltonian}
\end{align}

\noindent where $\Delta_\mathrm{AC}$ is the qubit frequency shift, $\alpha$ is qubit anharmonicity, $\Omega$ is the Rabi frequency of the drive, and $\delta = \omega_\mathrm{D} - \omega_\mathrm{Q}$ is the drive-qubit detuning. AC Stark tones cause measurement-induced qubit dephasing. For a shot-noise limited drive, this excess dephasing increases with higher Rabi frequencies ($\Gamma_{\varphi,\mathrm{AC}}\propto \Omega$) and reduced qubit-drive detunings ($\Gamma_{\varphi,\mathrm{AC}}\propto 1/\delta^2$) \cite{schuster2005ac,gambetta2006qubit}. Given a fixed AC Stark shift, to minimally dephase the qubit, we detune the microwave tone as much as possible from the qubit frequency. Stark tones are applied at $\approx4.1$ GHz; frequencies are not increased further to avoid populating readout resonators ($\approx4.3-4.4$ GHz). 

\subsubsection{Analysis of decoherence sources}

\begin{figure}[htbp]
\centering
\includegraphics[width=0.9\linewidth]{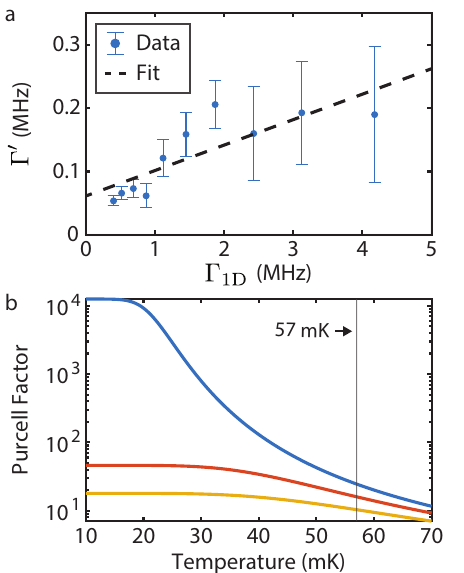}
\caption{\textbf{Decoherence sources.} (a) Waveguide temperature. The waveguide thermal occupation is obtained by varying $\Gamma_\mathrm{1D}$ of qubit 1. For each setting of $\Gamma_\mathrm{1D}$, $\Gamma'$ is obtained from a fit to the qubit transmission. Measured values of $\Gamma'$ are shown in blue, with error bars denoting $95\%$ confidence intervals. The dotted black line shows the fit of $\Gamma'$ and $\Gamma_\mathrm{1D}$ according to \cref{eq:wgtemp}, yielding waveguide temperature of $T_\mathrm{WG} = 57.0^{+7.0}_{-9.6}$ mK. (b) Projected Purcell factors vs. waveguide temperature. The blue curve predicts Purcell factors given qubit $T_1 = 500$ $\mu s$, $T_2^* = 500$ $\mu s$ and $\Gamma_\mathrm{1D}^f=4$ MHz. The red curve is calculated using average measured qubit $T_{1,\mathrm{avg}}=13.7$ $\mu s$, $T_{2,\mathrm{avg}}^*=8.3$ $\mu s$, and $\Gamma_\mathrm{1D}^f=2$ MHz. The yellow curve is obtained using average measured qubit lifetimes, $\Gamma_\mathrm{1D}^f=2$ MHz, and calculated qubit broadening.}
\label{fig:wgtemp}
\end{figure}

Observable Purcell factors are limited by several sources, including the intrinsic decoherence of the qubits, imperfect chirality, the finite thermal occupation of the waveguide, and parametric drive related effects. Having already discussed intrinsic decoherence (\cref{appendix:Device:summary}) and imperfect chirality (\cref{appendix:Device:chirality tuneup}), we turn our attention towards the thermal occupation of the waveguide and effects from parametric drives.

Due to imperfect thermalization, the waveguide contains a finite thermal population, which results in an non-zero effective waveguide temperature. The waveguide temperature is obtained by measuring the intrinsic decoherence ($\Gamma'$) while varying the waveguide decay rate ($\Gamma_\mathrm{1D}$) of emitter qubit 1 \cite{joshi2023resonance}. We drive a single coupler, resulting in a bidirectional qubit-waveguide coupling, and sweep the coupler pump power to vary $\Gamma_\mathrm{1D}$. For a finite thermal occupation in the waveguide, we expect excess intrinsic decoherence as the waveguide decay rate increases. In this case, by following a master equation treatment \cite{mirhosseini2019cavity}, the total qubit decay and decoherence are enhanced as $\Gamma_\mathrm{1}^\mathrm{th} = (2n_\mathrm{th}+1)\Gamma_\mathrm{1D} + \Gamma_0'$ and $\Gamma^\mathrm{th}_2 = \Gamma_1^\mathrm{th}/2 + \Gamma_\varphi$, where $\Gamma_1^\mathrm{th}$ and $\Gamma_2^\mathrm{th}$ are the enhanced decay and decoherence rates, $n_\mathrm{th}$ is the thermal occupation of the waveguide, $\Gamma_0'$ is the qubit internal decay rate, and $\Gamma_\varphi$ is the pure dephasing. In this treatment, the qubit internal decay and pure dephasing do not arise from qubit-waveguide coupling and are unaffected by a finite waveguide temperature. The intrinsic decoherence at finite temperature is given by $\Gamma' = 2\Gamma_2^\mathrm{th}-\Gamma_\mathrm{1D}$. Substituting the thermally enhanced decoherence and decay rates gives 
\begin{equation}
\Gamma' = 2 n_\mathrm{th}\Gamma_\mathrm{1D} + (\Gamma_0' + 2\Gamma_\varphi)
\label{eq:wgtemp}
\end{equation}

By fitting $\Gamma'$ and $\Gamma_\mathrm{1D}$ to \cref{eq:wgtemp}, we obtain the waveguide's thermal occupation and effective temperature $T_\mathrm{WG}$ using the Bose-Einstein distribution ($n_\mathrm{th} = 1/[e^{\hbar\omega/k_bT_\mathrm{WG}}-1]$, where $\hbar$ is the reduced Planck constant, $\omega$ is the qubit frequency, and $k_b$ is the Boltzmann constant). The measured data and fit is shown in \cref{fig:wgtemp}a, giving $T_\mathrm{WG} = 57.0^{+7.0}_{-9.6}$ mK. Using the measured waveguide temperature, we can then assess the limitation on Purcell factor. Averaging across the qubits, we use a qubit $T_{1,\mathrm{avg}}= 13.7$ $\mu s$ and $T^*_{2,\mathrm{avg}}= 8.3$ $\mu s$ to predict Purcell factors for $T_\mathrm{WG}=57$ mK and $\Gamma_\mathrm{1D}=2$ MHz. This yields $P_\mathrm{1D} \approx 16$, which agrees reasonably well with the best observed Purcell factors (see \cref{tab:table1}). 

For the best performing emitter qubits, we deduce that excess decoherence arising from parametric drives is minimal. Parametric pumps can increase decoherence via (1) decay through un-desired sidebands or (2) by exciting the qubits parasitically. (1) Sideband decay is ideally suppressed by the CCA waveguide. This is confirmed by comparing the temperature measurement to independent decoherence measurements. The intercept of the temperature fit (\cref{fig:wgtemp}a), corresponding to intrinsic decoherence at $\Gamma_\mathrm{1D}=0$, is $\Gamma_0' + 2\Gamma_\varphi = 61.5\pm45.4$ kHz. This agrees well with the $T_2^*$ measurement of qubit 1 in the absence of waveguide coupling, which yields $2\Gamma_2 = 68$ kHz ($\Gamma_2 = 1/2 \pi T_2^*$). The agreement between $2\Gamma_2$ (obtained from a Ramsey measurement) and $\Gamma'_0 + 2\Gamma_\varphi$ (obtained from waveguide temperature measurements) indicates that sideband decay is negligible, in contrast to previous work \cite{joshi2023resonance}. In prior work, $\Gamma'_0 + 2\Gamma_\varphi$ obtained from waveguide temperature measurements exceeded a measurement of $2\Gamma_2$ by $\approx 200$ kHz. (2) With respect to decoherence arising from parasitic qubit driving, we comment that the measured waveguide temperature is in agreement with previous experiments using the same measurement setup \cite{shah2024stabilizing}. This indicates that the extraction of waveguide temperature is still accurate in the presence of varying parametric pump powers. Therefore, for the best performing qubits, the parametric drives are not significantly increasing the decoherence ($\Gamma'$).

Not all qubits saturate the waveguide-temperature-imposed limit on Purcell factor. We attribute this to parasitic qubit driving, which can vary for each individual emitter. This effect arises due to a small but finite transverse coupling between flux lines and coupler elements, which reaches $\approx100$ Hz in electromagnetic simulations (Sonnet \textregistered). Because of this transverse interaction, flux pumps can populate couplers, which in turn coherently drives emitter qubits. This coherent driving, when combined with the frequency conversion that arises in the presence of flux pumping, can result in emitter saturation on the order of tens of kHz. For a typical on-chip parametric drive power of -40 dBm and 100 Hz of coupler to flux-line coupling, this effect can reduce Purcell factors to $\approx 9$ (using the same qubit and waveguide parameters as above). We expect to eliminate the coupler to flux-line transverse coupling by engineering the electromagnetic environment; this will involve adjusting flux-line geometries and increasing metal shielding.

Purcell factors may be improved in a straightforward fashion by reducing waveguide temperatures, which have been reported to be as low as 35 mK \cite{scigliuzzo2020primary}. \cref{fig:wgtemp}b shows projected Purcell factors vs. waveguide temperature for observed $T_{1,\mathrm{avg}}= 13.7$ $\mu s$ and $T^*_{2,\mathrm{avg}}= 8.3$ $\mu s$ with (yellow) and without (red) the inclusion of flux-line to coupler transverse coupling. For future device iterations, increasing the emitter qubit frequency exponentially reduces decoherence arising from the finite waveguide temperature.  Improving intrinsic qubit decoherence is another approach towards improving Purcell factors. The blue curve in \cref{eq:wgtemp}b shows projected Purcell factor vs. waveguide temperature for $T_1=500$ $\mu s$ and $T_2^*=500$ $\mu s$, which should be achievable for transmon qubits \cite{bland2025millisecondtransmon}. In this projection, using a $\Gamma_\mathrm{1D}^f=4$ MHz, improvements in waveguide temperature and qubit intrinsic decoherence bring Purcell factors above $10^4$. 

\subsection{Inter-chip loss}
\label{appendix:Device:waveguide}

The two chips are connected by a 45 cm superconducting aluminum cable. We obtain the photon loss between the chips by preparing and emitting the state $(|0\rangle+|1\rangle)/\sqrt{2}$ from qubits 2 and 3 in independent experiments, as done in \cite{kurpiers2018deterministic}. We measure the emitted field in each experiment, and compare the square of the emitted field to obtain the loss $\zeta = \int |\langle \hat{a}^\mathrm{Q3}_\mathrm{out}(t)\rangle|^2 dt / \int |\langle \hat{a}^\mathrm{Q2}_\mathrm{out}(t)\rangle|^2 dt$. We note that the square of the averaged field is not in general equal to the total power emitted by a qubit $\int |\langle \hat{a}_\mathrm{out}(t)\rangle|^2 dt \neq \int \langle \hat{a}_\mathrm{out}^\dagger (t) \hat{a}_\mathrm{out} (t)\rangle dt$. However, for a known qubit state $\alpha|g\rangle+\beta|e\rangle$ (where $\alpha,\beta\neq0$), the emitted field squared is proportional to the total power. This allows us to determine the photon loss by measuring only the emitted field. We obtain inter-chip photon loss of $\eta^2 = 17.2\%$, in line with previously reported values \cite{kurpiers2018deterministic, almanakly2025deterministic}. We attribute inter-chip loss to connector loss and loss in printed-circuit board (PCB) traces. The aluminum cable is connected to each device via a PCB with MMPX connectors, and PCB traces are made of gold-plated copper on a dielectric substrate. These loss sources can be reduced by directly wire-bonding superconducting cables to device chips \cite{niu2023low}.

\section{Device Design}
\label{appendix:Design}

\subsection{Chiral qubit-waveguide interaction}
\label{appendix:Design:chiral}

\begin{figure*}[t!]
\centering
\includegraphics[width=1\linewidth]{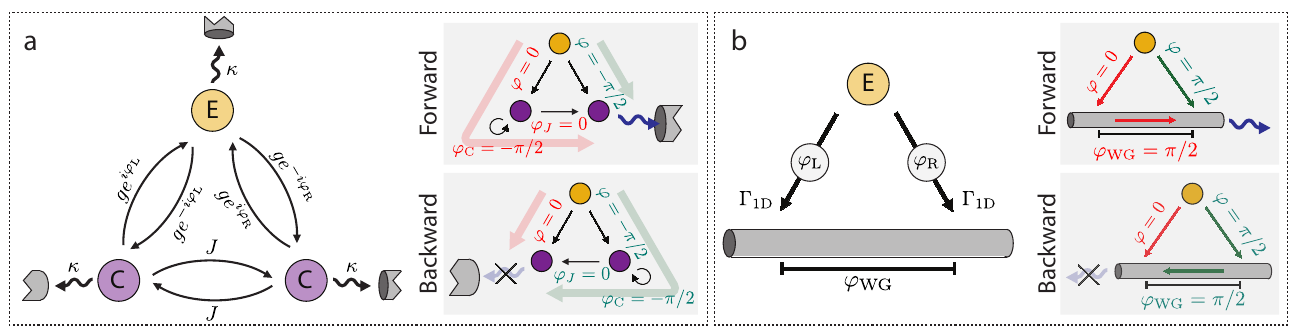}
\caption{\textbf{Chiral qubit-waveguide interaction.} (a) Three-mode Hamiltonian picture, where chirality arises from interference between photons transport in the loop. Inset: Interference pathways for forward and backward emission. (b) Giant atom picture, where chirality arises from interference between photons emitted at spatially separated points. Inset: Interference pathways for forward and backward emission.}
\label{fig:chiral interaction}
\end{figure*}

The chiral qubit-waveguide interaction is interference-based and can be modeled using two distinct but equivalent pictures. Here, we present a simplified explanation of both approaches. The first framework, which is presented in the \cref{fig:Fig1}b, consists of treating all emitter and CCA modes individually, with coherent couplings between modes. A minimal toy model for chirality can be obtained using just three modes - the emitter and two couplers. This model is depicted in \cref{fig:chiral interaction}a, with the emitter shown in yellow and couplers shown in purple. For simplicity, we assign a dissipation rate $\kappa$ to each mode and treat each mode bosonically. Couplers interact with hopping rate $J$. The emitter interacts with the left and right couplers with rates $g e^{i\varphi_\mathrm{L}}$ and $g e^{i\varphi_\mathrm{R}}$, respectively (The $g$ used here is interchangeable with $g_\mathrm{eff}$ used in the main text). In this loop, the combination of a synthetic gauge field (obtained by tuning $\varphi_\mathrm{L}$ and $\varphi_\mathrm{R}$) and on-site dissipations $\kappa$ enables directional transport \cite{clerk2022introduction}. 

Directionality arises from interference between the possible paths a photon can take from one site to another. For example, a photon traversing from the emitter to the left coupler can take two paths: (1) emitter $\rightarrow$ left coupler, or (2) emitter $\rightarrow$ right coupler $\rightarrow$ left coupler. To create a forward chiral qubit, the two paths from emitter to left coupler should destructively interfere, while the equivalent paths to the right coupler should constructively interfere. This can be done by tuning the relative phase $\Delta\varphi = \varphi_\textrm{R} - \varphi_\textrm{L}$ to $\pi/2$ (see \cref{fig:Fig1}b insets). In this case, a photon picks up a phase of -$\varphi_\textrm{L}$ in path (1). In path (2), emitter-right coupler hopping gives a phase of -$\varphi_\textrm{R}$, and an additional -$\pi/2$ is acquired from the phase response of the right coupler in the presence of dissipation. We denote this coupler phase as $\varphi_\mathrm{C}$. In path (2), the hopping phase from the right coupler to left coupler is $\varphi_J=0$. Considering interference paths terminating at the left (right) coupler, the relative phase between the paths is $\Delta\varphi + \pi/2 = \pi$ $(\Delta\varphi - \pi/2 = 0)$, giving destructive (constructive) interference. These interference paths are shown for forward and backward photon transport in the \cref{fig:chiral interaction}a inset.

We clarify this picture more formally by starting with the three-mode Hamiltonian of \cref{fig:chiral interaction}a.

\begin{equation}
\begin{split}
\hat{H}/\hbar &= \omega_0 \hat{a}_\mathrm{E}^\dagger \hat{a}_\mathrm{E} + \omega_0 \hat{a}_\mathrm{L}^\dagger \hat{a}_\mathrm{L} + \omega_0 \hat{a}_\mathrm{R}^\dagger \hat{a}_\mathrm{R} \\ 
& +g(e^{i\varphi_\mathrm{L}}\hat{a}^\dagger_\mathrm{E} \hat{a}_\mathrm{L} + e^{-i\varphi_\mathrm{L}}\hat{a}_\mathrm{E} \hat{a}^\dagger_\mathrm{L}) \\
& +g(e^{i\varphi_\mathrm{R}}\hat{a}^\dagger_\mathrm{E}\hat{a}_\mathrm{R} + e^{-i\varphi_\mathrm{R}}\hat{a}_\mathrm{E} \hat{a}^\dagger_\mathrm{R}) \\
& +J(\hat{a}^\dagger_\mathrm{L} \hat{a}_\mathrm{R} + \hat{a}_\mathrm{L} \hat{a}^\dagger_\mathrm{R})
\end{split}
\end{equation}

\noindent We set the frequencies of all modes to $\omega_0$. The raising (lowering) operators for the emitter, left coupler, and right coupler are $\hat{a}^\dagger_\mathrm{E}$ $(\hat{a}_\mathrm{E})$, $\hat{a}^\dagger_\mathrm{L}$ $(\hat{a}_\mathrm{L})$, $\hat{a}^\dagger_\mathrm{R}$ $(\hat{a}_\mathrm{R})$. The equations of motion in the frequency domain are then given as

\begin{equation}
-i\omega\hat{a}_\mathrm{E} = -(i\omega_0+\frac{\kappa}{2})\hat{a}_\mathrm{E}-ige^{i\varphi_\mathrm{L}}\hat{a}_\mathrm{L}-ige^{i\varphi_\mathrm{R}} \hat{a}_\mathrm{R} -\sqrt{\kappa}\hat{a}_\mathrm{E,in}
\end{equation}

\begin{equation}
-i\omega\hat{a}_\mathrm{L} = -(i\omega_0+\frac{\kappa}{2})\hat{a}_\mathrm{L} -iJ\hat{a}_\mathrm{R} - ige^{-i\varphi_\mathrm{L}}\hat{a}_\mathrm{E} - \sqrt{\kappa}\hat{a}_\mathrm{L,in} 
\end{equation}

\begin{equation}
-i\omega\hat{a}_\mathrm{R} = -(i\omega_0+\frac{\kappa}{2})\hat{a}_\mathrm{R} -iJ\hat{a}_\mathrm{L} - ige^{-i\varphi_\mathrm{R}}\hat{a}_\mathrm{E} - \sqrt{\kappa}\hat{a}_\mathrm{R,in} 
\end{equation}

\noindent The input-output relations are 

\begin{equation}
\hat{a}_{j,\mathrm{out}} = \hat{a}_{j,\mathrm{in}} + \sqrt{\kappa}\hat{a}_j
\end{equation}

\noindent where $\hat{a}_{j,\mathrm{in (out)}}$ is the input (output) mode amplitude of the mode $j$ in units of $\sqrt{\mathrm{Hz}}$ ($j=\mathrm{E,L,R}$). We can use these relations to derive the scattering matrix $\mathbf{S}$, where $\mathbf{\hat{a}_\mathrm{out}}$ = $\mathbf{S\hat{a}_\mathrm{in}}$ ($\mathbf{\hat{a}_\mathrm{out}} = [\hat{a}_{\mathrm{E,out}}, \hat{a}_{\mathrm{L,out}}, \hat{a}_{\mathrm{R,out}},]^\intercal$, $\mathbf{\hat{a}_\mathrm{in}} = [\hat{a}_{\mathrm{E,in}}, \hat{a}_{\mathrm{L,in}}, \hat{a}_{\mathrm{R,in}},]^\intercal$). The scattering matrix elements are given as 

\begin{equation}
    [\mathbf{S}]_{j,k} = \delta_{j,k} - i\kappa [\mathbf{G}]_{j,k}
\end{equation}

\noindent where $j$ and $k$ represent mode indices and $[\mathbf{G}]_{j,k}$ is the Green's function representing the amplitude of mode $j$ arising from exciting mode $k$. The Green's function is defined as $[\mathbf{G}]_{j,k} = \left[\mathbf{H}^{-1}_\mathrm{eff}\right]_{j,k}$, where

\begin{equation}
\mathbf{H_{eff}} 
= \begin{bmatrix} 
\delta +i\kappa/2 & -ge^{i\varphi_\mathrm{L}} & -ge^{i\varphi_\mathrm{R}} \\  
-ge^{-i\varphi_\mathrm{L}} & \delta + i\kappa/2 & -J \\
-ge^{-i\varphi_\mathrm{R}} & -J & \delta +i\kappa/2\\
\end{bmatrix} 
\end{equation}

\noindent and $\delta = \omega - \omega_0$. We then consider the amplitude at the output ports of coupler modes L and R when exciting the emitter. For a fully chiral qubit, we expect $[\mathbf{S}]_\mathrm{R,E} \neq 0$ and $[\mathbf{S}]_\mathrm{L,E} = 0$. This condition is equivalent to $[\mathbf{G}]_\mathrm{R,E} \neq 0$ and $[\mathbf{G}]_\mathrm{L,E} = 0$. Taking $\delta=0$, $\Delta\varphi = \varphi_\mathrm{R} -\varphi_\mathrm{L} = \pi/2$, and solving for the Green's functions, we find 

\begin{equation}
    [\mathbf{G}]_\mathrm{R,E} = \frac{4ig(2J+\kappa)}{\kappa(\kappa^2+4J^2+8g^2)}
\end{equation}

\begin{equation}
\label{eq:green fcn left}
    [\mathbf{G}]_\mathrm{L,E} = \frac{4g(2J-\kappa)}{\kappa(\kappa^2+4J^2+8g^2)}
\end{equation}

\noindent We see that for $\kappa=2J$, $[\mathbf{G}]_\mathrm{L,E} = 0$, meaning that there is no photon transport from the emitter mode to the left coupler. There is destructive interference between the two terms in the numerator of $[\mathbf{G}]_\mathrm{L,E}$. As discussed in \cite{clerk2022introduction}, the Green's functions can be expanded directly, with individual terms representing different photon transport paths. For example, consider the lowest order terms in $J$ and $g$ for $[\mathbf{G}]_\mathrm{L,E}$, provided $J, g \ll \kappa$. This gives 

\begin{equation}
\label{eq:green expanded}
    [\mathbf{G}]_\mathrm{L,E} \approx -\frac{4g}{\kappa^2} + \frac{8gJ}{\kappa^3} + ...
\end{equation}

\noindent The first two terms given represent the two photon transport paths shown in the left inset of \cref{fig:Fig1}b and discussed previously. The first term describes a photon moving from the emitter directly to the left coupler. The second term describes transport from the emitter, to the right coupler, to the left coupler. We note that a photon traversing a mode experiences the un-coupled Green's function of that mode (Green's function in the absence of all other modes). For $\delta=0$, this is $\frac{1}{i\kappa/2}$. This allows us to break down each transport path explicitly, as given below.

\begin{equation}
\mathrm{Path_{E\rightarrow L}} = -\frac{4g}{\kappa^2} = \left( \frac{1}{i\kappa/2} \right)g\left( \frac{1}{i\kappa/2} \right)  
\end{equation}

\begin{equation}
\mathrm{Path_{E\rightarrow R \rightarrow L}} = \frac{8gJ}{\kappa^3} =\left( \frac{1}{i\kappa/2} \right)(-ig)\left( \frac{1}{i\kappa/2} \right) J \left( \frac{1}{i\kappa/2} \right)
\end{equation}

\noindent In each path, a photon picks up the hopping phase between modes and the phase from the un-coupled Green's functions of each mode. The Green's function given in \cref{eq:green fcn left} unambiguously shows full destructive interference. However, using the expansion given in \cref{eq:green expanded} allows a more physical picture by listing out the possible paths a photon may take. The higher order terms in \cref{eq:green expanded} represent higher order transport paths. One concrete example is emitter $\rightarrow$ right coupler $\rightarrow$ emitter $\rightarrow$ right coupler $\rightarrow$ left coupler.

The second framework we consider is a waveguide picture of the emission, as depicted in \cref{fig:chiral interaction}b. In this interpretation, the emitter qubit can be considered as a giant atom \cite{Kockum2018Apr} coupling to a waveguide at two spatially separated points, where the two points are separated by a propagation phase $\varphi_\mathrm{WG}$. For our emitter qubit, this propagation phase can be obtained using the dispersion relation of the CCA shown in \cref{fig:passband}a. The qubit emits at the two coupling points with phases $\varphi_\mathrm{L}$ and $\varphi_\mathrm{R}$. Again, chirality emerges due to interference. For $\varphi_\mathrm{WG} = \pi/2$ and $\Delta\varphi = \varphi_\mathrm{R} - \varphi_\mathrm{L} = \pi/2$, the qubit emits directionally in the forward direction. Considering the backward direction, emission from the left (right) coupling has phase $\varphi_\mathrm{L}$ ($\varphi_\mathrm{R}$), and the right emission picks up phase $\varphi_\mathrm{WG}$ while propagating to the left point. The relative phase between the two emissions is then $\Delta \varphi + \varphi_\mathrm{WG} = \pi$, resulting in destructive interference. These interference paths are depicted in the \cref{fig:chiral interaction}b inset for forward and backward emission. The input-ouput relations for this picture are given below. 

\begin{equation}
\hat{a}^f_\mathrm{out} = \hat{a}^f_\mathrm{in} + (1+e^{i(\Delta\varphi-\varphi_\mathrm{WG})}) \sqrt{\frac{\Gamma_\mathrm{1D}}{2}}\hat{\sigma}^-
\end{equation}

\begin{equation}
\hat{a}^b_\mathrm{out} = \hat{a}^b_\mathrm{in} + (1+e^{i(\Delta\varphi+\varphi_\mathrm{WG})}) \sqrt{\frac{\Gamma_\mathrm{1D}}{2}}\hat{\sigma}^-
\end{equation}

\noindent and can be used to derive the transmission for a chiral qubit (\cref{eq:transmission}) using the SLH formalism \cite{joshi2023resonance}. Here, $\hat{a}^{f(b)}_\mathrm{in}$ is the forward (backward) input mode and $\hat{a}^{f(b)}_\mathrm{out}$ is the forward (backward) output mode. $\Gamma_\mathrm{1D}$ is the decay rate at each coupling point.  

\subsection{Circuit design}
\label{appendix:Design:circuit}

\subsubsection{Coupled cavity array}

\begin{figure*}[t!]
\centering
\includegraphics[width=1\linewidth]{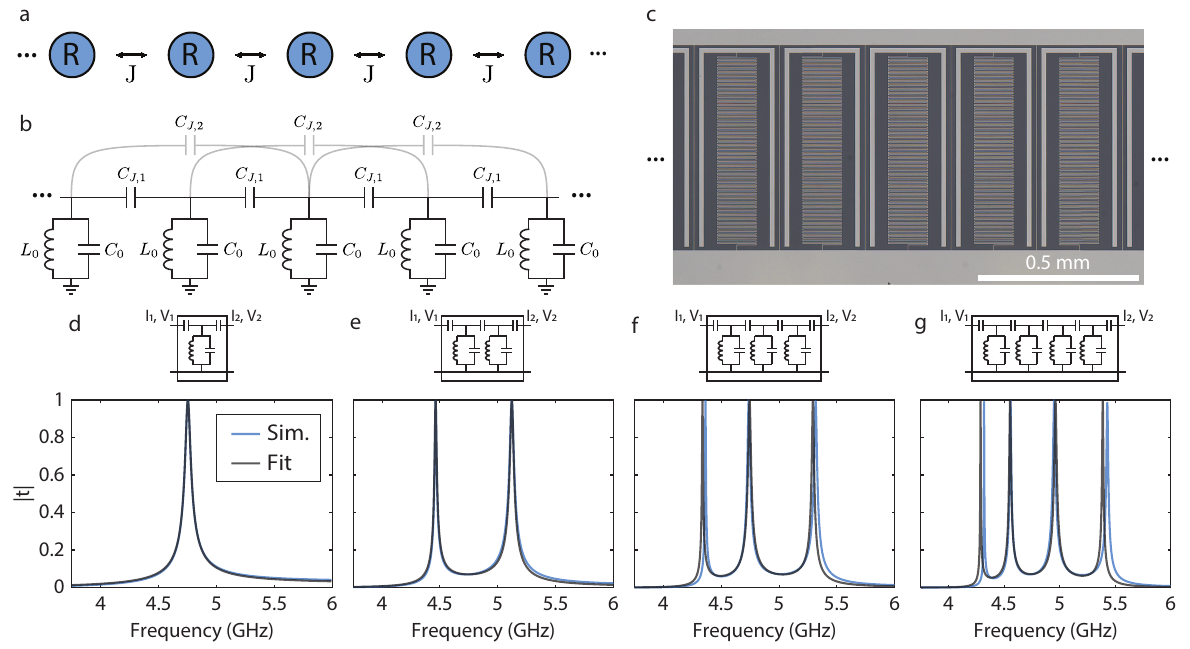}
\caption{\textbf{Circuit model of linear coupled cavity array.} (a) Schematic of tight-binding chain model for the coupled cavity array waveguide, showing resonators (blue) coupled in a chain with hopping amplitude $J$. (b) Circuit diagram of linear CCA, showing LC oscillators coupled to nearest neighbors by capacitance $C_{J,1}$. Parasitic capacitance between next-nearest neighbors is shown by capacitance $C_{J,2}$. (c) Optical image of a section of the CCA, showing a chain of lumped element resonators, with each resonator containing an inductive meander and capacitive claw. 2-port network for (d) one, (e) two, (f) three, and (g) four oscillator chains capacitively coupled to input-output ports. Below each network diagram (d,e,f,g) is the simulated transmission profile for the network (using the final unit cell design) and the transmission fit, derived using ABCD matrices.}
\label{fig:linearMM}
\end{figure*}

The CCA waveguide consists of a tight-binding chain of coupled resonators, as shown schematically in \cref{fig:linearMM}a. To construct the CCA, we capacitively couple lumped-element LC resonators, as shown in \cref{fig:linearMM}b. Each resonator has a self-inductance $L_0$ and bare self-capacitance $C_0$. Adjacent resonators are coupled by a capacitance $C_{J,1}$, and next-nearest neighbor resonators may be coupled by parasitic capacitance $C_{J,2}$. This circuit is equivalent to a multi-pole filter and has been analyzed in previous works \cite{ferreira2021collapse, zhang2023superconducting}. For completion, we reproduce previous analyses below. We write the Hamiltonian of the circuit as 

\begin{equation}
\label{eq:Hexact}
H = \frac{1}{2}\mathbf{Q}^\intercal\mathbf{C}^{-1}\mathbf{Q} + \frac{1}{2L_0}\mathbf{\Phi}^\intercal\mathbf{\Phi}.
\end{equation}

\noindent Here, $\mathbf{Q} = [Q_1,Q_2 ,...]^\intercal$ and $\mathbf{\Phi} = [\Phi_1,\Phi_2 ,...]^\intercal$, where $Q_j$ and $\Phi_j$ are the charge and flux variables at node $j$ (located on resonator $j$). The flux variable is defined as $\Phi_j(t) = \int_{-\infty}^tV_j(t')dt'$, where $V_j$ is the voltage at node $j$. The capacitance matrix is given as 

\begin{equation}
\mathbf{C} =  C_\mathrm{tot}\mathbbm{1} - C_{J,1}\mathbf{J}_1-C_{J,2}\mathbf{J}_2
\end{equation}

\noindent where $C_\mathrm{tot} = C_0 + 2C_{J,1} + 2C_{J,2}$ and $[\mathbf{J}_j]_{k,l} = \delta_{k,j+l} + \delta_{j+k,l}$. Assuming no parasitic next-nearest neighbor capacitance ($C_{J,2} = 0$) and small coupling capacitance ($C_{J,1} \ll C_\mathrm{tot}$), the inverse of the capacitance matrix can be calculated to first order as shown below.

\begin{equation}
\begin{split}
\mathbf{C}^{-1} &= \left[C_\mathrm{tot} \left(\mathbbm{1} - \frac{C_{J,1}}{C_\mathrm{tot}} \mathbf{J}_1 \right) \right]^{-1} \\
&=\frac{1}{C_\mathrm{tot}}\left[ \mathbbm{1} + \frac{C_{J,1}}{C_\mathrm{tot}}\mathbf{J}_1 + \mathcal{O} \left( \left(\frac{C_{J,1}}{C_\mathrm{tot}}\right)^2 \right) \right ] \\
&\approx \frac{1}{C_\mathrm{tot}}\mathbbm{1} + \frac{C_{J,1}}{C_\mathrm{tot}^2}\mathbf{J_1}
\end{split}
\end{equation}

\noindent The Hamiltonian can then be written as 

\begin{equation}
\label{eq:Hcircuit}
H \approx \sum_{j} \frac{Q^2_j}{2C_\mathrm{tot}} + \frac{\Phi_j^2}{2L_0} + \frac{C_{J,1}}{C_\mathrm{tot}^2} Q_j Q_{j+1}
\end{equation}

\noindent To first order, there is only coupling between adjacent resonators, but expanding the capacitance matrix to higher orders results in long-range coupling beyond nearest neighbor resonators, even in the absence of direct capacitive coupling beyond nearest neighbors ($C_{J,2}=0$). We can re-write the Hamiltonian using raising (lowering) operators $\hat{a}_j^\dagger$ ($\hat{a}_j$) using the following definitions.

\begin{equation}
\label{eq:Phi}
\hat{\Phi}_j = \sqrt{\frac{\hbar Z_0}{2}}(\hat{a}_j + \hat{a}_j^\dagger)
\end{equation}

\begin{equation}
\label{eq:Q}
\hat{Q}_j = -i\sqrt{\frac{\hbar}{2 Z_0}}(\hat{a}_j - \hat{a}_j^\dagger)
\end{equation}

\noindent where $\Phi_j$ and $Q_j$ are replaced by their respective quantum operators ($\hat{\Phi}_j$ and $\hat{Q}_j$) and $Z_0 = \sqrt{L_0/C_\mathrm{tot}}$. Raising and lowering operators satisfy canonical commutation relations $[\hat{a}_j, \hat{a}_k^\dagger] = \delta_{j,k}$. Substituting \cref{eq:Phi}, \cref{eq:Q} into \cref{eq:Hcircuit} yields the tight-binding Hamiltonian as described in the main text.

\begin{equation}
\label{eq:MM Hamiltonian}
\hat{H}/\hbar =  \sum_j \omega_0 \hat{a}_j^\dagger \hat{a}_j +
J(\hat{a}_j^{\dagger} \hat{a}_{j+1} + \hat{a}_j \hat{a}_{j+1}^\dagger)
\end{equation}

\noindent Here $\omega_0 = 1/\sqrt{L_0 C_\mathrm{tot}}$ is the resonator frequency and $J = \frac{C_{J,1}}{2C_\mathrm{tot}}\omega_0$ is the hopping amplitude. Taking a plane wave ansatz, the dispersion relation of this Hamiltonian can be found to be $\omega(k) = \omega_0 + 2J\cos(kd)$, where $k$ is the wave vector and $d$ is the lattice constant. In our device, $C_{J,1}/C_\mathrm{tot} \approx 0.14$, so we expect the passband to deviate from the ideal tight-binding picture. In this case, by solving the exact Hamiltonian \cref{eq:Hexact}, the exact dispersion relation can be obtained \cite{ferreira2021collapse}, given by

\begin{equation}
\omega(k) = \frac{\omega_{0,\mathrm{bare}}}{\sqrt{1+\frac{4C_{J,1}}{C_0}\sin^2(kd/2)}}
\end{equation}

\noindent where $\omega_{0,\mathrm{bare}} = 1/\sqrt{L_0C_0}$ is the bare resonator frequency in the absence of coupling capacitances.

An optical image of the coupled resonator waveguide is shown in \cref{fig:linearMM}c. Each resonator contains a 2 $\mu$m wide meandered inductor ($L_0$) and a capacitive claw. The capacitive claw accounts for the resonator's bare self-capacitance ($C_0$) and coupling capacitance to neighboring resonators ($C_{J,1}$). We use electromagnetics simulations (Sonnet \textregistered) to extract circuit parameters for a given device geometry. We target design parameters of $L_0 = 6.5$ nH, $C_0 = 125$ fF, and $C_{J,1}= 24$ fF, which results in a passband of $\approx 4.2-5.6$ GHz. The measured passband does not fully agree with targeted values, which we attribute to slight differences in the simulated vs. actual box geometries affecting resonator self-capacitances.

We perform a series of simulations to extract circuit parameters. Given a proposed unit cell design, we perform the following steps:

\begin{enumerate}[label=\roman*.]
\item \emph{Extract $L_0$}: simulate a single resonator unit cell with inductive meander and capacitive claw, adding an ideal capacitor ($C_\mathrm{ideal}$) between the capacitive claw and the ground. Sweep the value of the ideal capacitor, and for each ideal capacitance, obtain the resonance frequency of the oscillator. Extract the inductance $L_0$ using the relation $L_0 C_\mathrm{tot} +L_0 C_\mathrm{ideal}= 1/\omega_0^2$, where $L_0$ is the slope of $1/\omega_0^2$ against $C_\mathrm{ideal}$.

\item \emph{Extract $C_0$ and $C_{J,1}$}: We use the ABCD matrix (transmission matrix) formalism to obtain $C_0$ and $C_{J,1}$ \cite{pozar2011microwave}. Given a two-port network (containing ports 1 and 2), the ABCD matrix is defined as 

\begin{equation}
\begin{bmatrix} V_1 \\ I_1 \end{bmatrix} = \begin{bmatrix} A & B \\ C & D  \end{bmatrix} \begin{bmatrix} V_2 \\ I_2 \end{bmatrix}
\label{eq:ABCD}
\end{equation}

\noindent where $V_1$, $V_2$ are voltages and $I_1$, $I_2$ are currents at ports 1 and 2, respectively. Examples of 2-port networks are given in the top of \cref{fig:linearMM}d,e,f,g. There is a general mapping between ABCD matrices and S-matrices \cite{pozar2011microwave}, which allows us to calculate S-parameters such as the transmission coefficient ($t$, or equivalently, $S_{21}$) given the ABCD matrix of an arbitrary circuit. The basic building block of the coupled cavity array is given in the top of \cref{fig:linearMM}d, and contains an LC oscillator with capacitive coupling to two input-output ports. To obtain $C_0$ and $C_{J,1}$, we first obtain the ABCD matrix for the circuit in \cref{fig:linearMM}d and map it to a transmission coefficient. We then perform a simulation of the single resonator, where the capacitive claws of adjacent resonators are included and used as the two input-output ports. This choice of ports allows for a faithful extraction of $C_{J,1}$. We then fit the simluated transmission coefficient $t$ between ports 1 and 2 using the derived transmission coefficient. The derived transmission coefficient $t(\omega,L_0,C_0,C_{J,1})$ is a function of the frequency $\omega$ and three input parameters $L_0$, $C_0$, $C_{J,1}$. To obtain capacitances, we use non-linear least squares fitting with variable inputs $C_0$ and $C_{J,1}$, keeping $L_0$ fixed. With the constraint on $L_0$, the self-capacitance $C_0$ determines the resonance frequency observed and the coupling capacitance $C_{J,1}$ determines the linewidth of the simulated mode. 

\item \emph{Confirm $C_{J,2}=0$}: The ABCD matrix allows for cascading of multiple sub-systems by simple matrix multiplication. For example, two cascaded two-port networks can be modeled as

\begin{align}
\begin{bmatrix} V_1 \\ I_1 \end{bmatrix} = \begin{bmatrix} A_1 & B_1 \\ C_1 & D_1  \end{bmatrix} \begin{bmatrix} A_2 & B_2 \\ C_2 & D_2 \end{bmatrix} \begin{bmatrix} V_2 \\ I_2 \end{bmatrix}.
\label{eq:ABCD_cascade}
\end{align}

\noindent This allows us to calculate ABCD matrices (and S-matrices) of multiple coupled resonators. We simulate two-, three-, and four-resonator chains with capacitively coupled input-output ports, shown in \cref{fig:linearMM}e, f, and g respectively. By fitting simulations to their respective transmission coefficients, we find consistent values for the extracted $C_0$ and $C_{J,1}$, indicating that next-nearest neighbor capacitances ($C_{J,2}$) are playing a minimal role.

\end{enumerate}

In our design, we include a thin metal bridge between each adjacent resonator to connect the ground plane above and below the circuit. These metal strips suppress slot-line modes, similar to air-bridges used in co-planar waveguides  \cite{Chen2014Feb}. We also include kinetic inductance in electromagnetic simulations. Because we use long meandered inductors, the kinetic inductance of aluminum can contribute to the total resonator self-inductance. From previous resonator tests where the width of inductive meanders are varied, we extract a kinetic inductance of aluminum films to be $39$ fH$/\square$. The kinetic inductance fraction of the CCA resonators, defined as $L_k/L_0$, is less than $5\%$. $L_k$ is the resonator inductance arising from kinetic inductance.

\subsubsection{Tapering section}

\begin{figure*}[t!]
\centering
\includegraphics[width=1\linewidth]{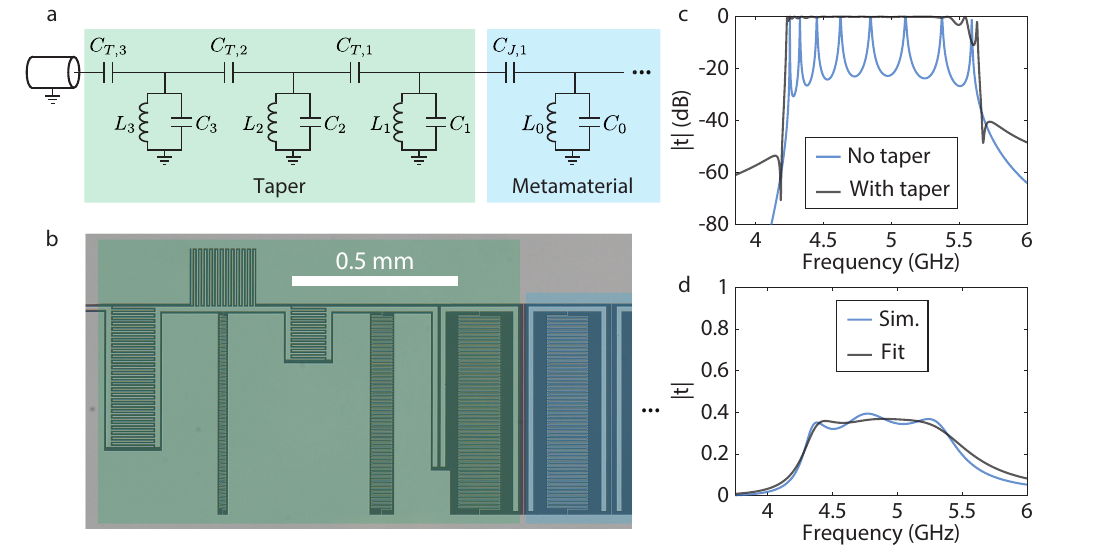}
\caption{\textbf{Tapering section.} (a) Circuit of tapering section (green) that impedance matches the linear coupled cavity array (blue) to a 50 $\Omega$ waveguide. The taper section contains three capacitively coupled LC oscillators. (b) Optical image of the taper circuit (shaded in green). Each taper resonator contains a capacitive claw and inductive meander. Inter-digitated capacitors are used to accommodate larger capacitance values in the taper. (c) Simulated transmission of the CCA waveguide in the absence and presence of tapers at the boundaries. Without a taper (blue), the transmission contains many ripples. With tapers included (black), the passband transmission is flat. (d) Simulated and fitted transmission response of the taper circuit containing the three taper resonators and a single CCA unit cell. }
\label{fig:taper}
\end{figure*}

Directly coupling the boundaries of a coupled-resonator chain to 50 $\Omega$ waveguides results in a transmission profile with large ripples and complicated dispersion. Because we desire a flat waveguide transmission profile in the passband for coupling qubits to the waveguide, we impedance match the resonator chain to the 50 $\Omega$ input-output ports. For this purpose, a tapering section is added to the chain boundary. Previous works have analyzed and implemented similar tapering sections for coupled-resonator waveguides \cite{sumetsky2003modeling, ferreira2021collapse, zhang2023superconducting}. The taper that we use contains three capacitively coupled LC resonators. The circuit diagram is shown in \cref{fig:taper}a, where the taper section is shaded in green and the periodic CCA is shaded in blue. An optical image of a CCA taper is shown in \cref{fig:taper}b, shaded in green. Similar to the CCA unit cell, each resonator in the taper contains an inductive meander and capacitive claw. The taper circuit contains nine free parameters: three self-inductances, three self-capacitances, and three coupling capacitances. We optimize over these parameters to maximize the transmission and minimize ripples in the waveguide transmission profile. The simulated transmission of the CCA waveguide in the presence and absence of a tapering section is shown in \cref{fig:taper}c, showing a flat passband when the taper is present. The final optimized taper parameters are $L_1 = 6.7$ nH, $C_1 = 107.5$ fF, $C_{T,1} = 35.0$ fF, $L_2 = 3.5$ nH, $C_2 = 174.7$ fF, $C_{T,2} = 103.8$ fF, $L_3 = 1.5$ nH, $C_3 = 331.4$ fF, $C_{T,3} = 436.4$ fF. To design the geometry of the taper circuit, we take the following steps:

\begin{enumerate}[label=\roman*.]

\item \emph{Extract inductances $L_1$, $L_2$, $L_3$}: simulate individual taper LC resonators with an ideal capacitor $C_\mathrm{ideal}$ in parallel with the capacitive claw. Repeating the procedure for extracting inductance $L_0$ of the CCA unit cell, we sweep $C_\mathrm{ideal}$ and obtain the resonance frequency $\omega_0$ for each simulation. $L_j$ $(j=1,2,3)$ can be obtained from $C_\mathrm{ideal}$ and $\omega_0$.

\item \emph{Extract $C_1$ and $C_{T,1}$}: perform a simulation containing the first taper resonator and the neighboring CCA unit cell. Couple the first output port to the taper resonator using a capacitor corresponding to $C_{T,1}$, and adjust the CCA unit cell so that its capacitive claw is used as the second output port. Fit the simulated transmission to the analytical transmission profile derived from ABCD matrices, taking $L_0$, $C_0$, $C_{J,1}$, and $L_1$, as fixed inputs. The outputs of the fit are $C_1$ and $C_{T,1}$.

\item \emph{Extract $C_2$ and $C_{T,2}$}: simulate the first and second taper resonators in a similar procedure to the previous step. The first output port of the simulation is capacitively coupled to the second taper resonator with capacitor $C_{T,2}$, and the second output port drives the capacitive claw of the first taper resonator, which has been extended to the boundary of the simulation. Fit simulated transmission to the analytical expression (derived from the ABCD formalism), taking $L_{1}$, $C_{1}$, $C_{T,1}$, and $L_{2}$ as inputs and outputting $C_{2}$ and $C_{T,2}$. After extracting the capacitancees of the first two taper resonators ($C_1$, $C_2$, $C_{T,1}$, $C_{T,2}$), we perform a simulation containing a CCA unit cell and the two taper cells. By comparing the simulated and expected transmission (where expected transmission is calculated using the extracted parameters), we confirm the extracted circuit values. 

\item \emph{Extract $C_3$ and $C_{T,3}$}: perform a simulation containing the entire taper section, including the three taper resonators and a single CCA unit cell. Similar to previous steps, the first output port is coupled to the outer-most taper resonator via capacitor $C_{T,3}$, and the second output port drives the capacitive claw of the CCA unit cell. We use the previously extracted parameters ($L_0$, $C_0$, $C_{J,1}$, $L_1$, $C_1$, $C_{T,1}$, $L_2$, $C_2$, $C_{T,2}$, $L_3$) as inputs to a fit of the simulated transmission and extract $C_3$ and $C_{T,3}$ from the fit. The result of this simulation and fit of the final taper is shown in \cref{fig:taper}d.
\end{enumerate}

\subsubsection{Emitters and Embedded couplers}

\begin{figure*}[t!]
\centering
\includegraphics[width=1\linewidth]{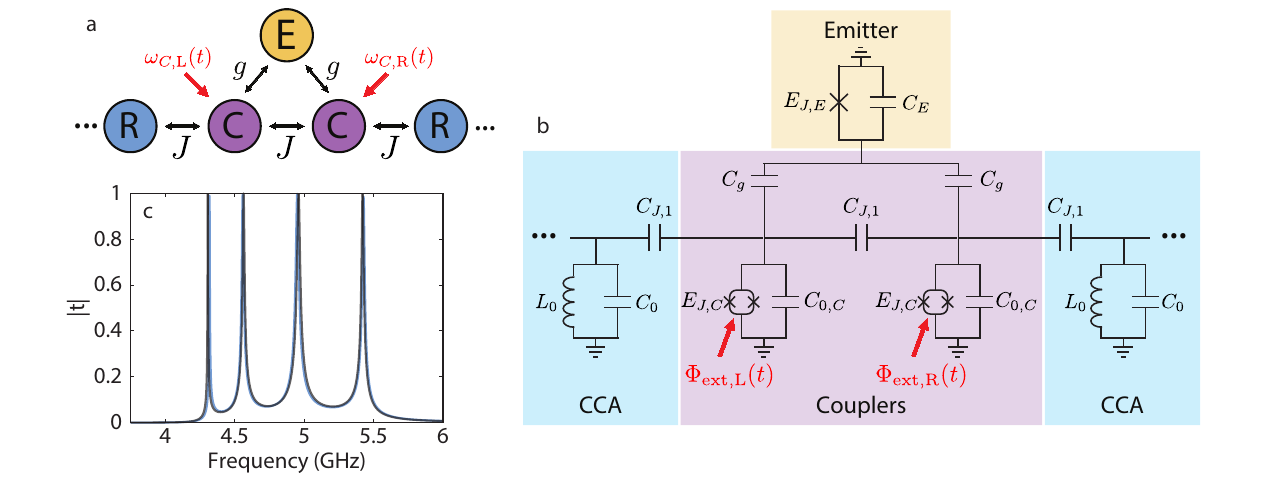}
\caption{\textbf{Emitters and embedded couplers} (a) Diagram of the emitter qubit (yellow) coupled to the coupled cavity array waveguide via two coupler modes (purple) embedded in the resonator chain (blue). (b) Circuit diagram of the emitter capacitively coupled to the CCA. (c) Simulated transmission response of four CCA modes capacitively coupled to input-output ports. The blue curve shows the response of four linear, lumped-element resonators (as in \cref{fig:linearMM}g) and the black curve shows the response of a chain of resonator-coupler-coupler-resonator.}
\label{fig:embedded coupler}
\end{figure*}

Each emitter qubit is coupled to two adjacent modes of the coupled cavity array, called couplers, as shown in \cref{fig:embedded coupler}a. The two couplers are embedded in the CCA, and the coupler-resonator and coupler-coupler interactions rates are both set to $J$, which is equal to the hopping amplitude in the CCA. The interaction rate between the emitter (shown in yellow) and a coupler (shown in purple) is $g$. We design the emitter qubit frequency to be in the waveguide bandgap and modulate the frequency of the coupler modes to convert the emitter qubit frequency into the passband, as shown in \cref{fig:embedded coupler}a. The equivalent circuit is shown in \cref{fig:embedded coupler}b. The emitter is a fixed-frequency qubit and the coupler modes are flux-tunable transmons. We tune coupler frequencies into resonance with linear CCA resonators via a DC flux threaded through the coupler SQUID loop. We frequency modulate couplers using a time-varying flux threading the SQUID loop. The Hamiltonian for a single emitter coupled to the CCA can be derived to first order following the same procedure as above, and the result is given below.
\begin{widetext}
\begin{equation}
\label{eq:Hqubit}
\begin{split}
H &\approx \frac{Q^2_E}{2C_{\mathrm{tot},E}} -E_{J,E}\cos \left(\frac{2\pi\Phi_{E}}{\Phi_0}\right) + \sum_{j=j_\mathrm{L},j_\mathrm{R}} \left( -E_{J,C}\cos \left(\frac{2\pi\Phi_{j}}{\Phi_0}\right) + \frac{C_g}{C_{\mathrm{tot},E} C_{\mathrm{tot},j}} Q_E Q_j \right) \\
& \sum_{j\neq j_\mathrm{L} j_\mathrm{R}} \frac{\Phi_j^2}{2L_0} + \sum_{j} \left( \frac{Q^2_j}{2C_{\mathrm{tot},j}} + \frac{C_{J,1}}{C_{\mathrm{tot},j} C_{\mathrm{tot},j+1}} Q_j Q_{j+1} \right) \\ 
\end{split}
\end{equation}
\end{widetext}

\noindent Here, $Q_E$ and $\Phi_E$ are the charge and flux variables of the emitter qubit. $C_{\mathrm{tot},E} = C_E + 2C_g$ describes the total capacitance of the emitter and $E_{J,E}$ is the emitter's Josephson energy. $\Phi_0 = h/2e$ is the magnetic flux quantum, where $h$ is Planck's constant and $e$ is the electron charge. Because the two couplers are immediate neighbors, $j_\mathrm{R}=j_\mathrm{L}+1$. $Q_j$ and $\Phi_j$ are the charge and flux of CCA mode $j$. $E_{J,C}$ is the coupler Josephson energy and $C_g$ is the emitter-coupler mutual capacitance. The total capacitances of CCA modes are  

\begin{equation}
C_{\mathrm{tot},j} = 
\begin{cases}
 C_{0,C} + 2C_{J,1} + C_g  & j= j_\mathrm{L}, j_\mathrm{R} \\
C_0+2C_{J,1}  &\mathrm{otherwise}
\end{cases}
\end{equation}

By promoting the charge and flux variables to their respective quantum operators (following equivalent relations to \cref{eq:Phi} and \cref{eq:Q}), we can re-write \cref{eq:Hqubit} using raising and lowering operators. Here, we Taylor expand the cosine Josephson potential for emitter and coupler qubits, keeping the first three terms as shown below (for the emitter).

\begin{equation}
\begin{split}
-E_{J,\mathrm{E}}\cos\left(\frac{2\pi\Phi_\mathrm{E}}{\Phi_0}\right) \approx -E_{J,\mathrm{E}}\left[1 - \frac{1}{2}\left(\frac{2 \pi \Phi_\mathrm{E}}{\Phi_0}\right)^2\right. \\
+ \left. \frac{1}{24}\left(\frac{2 \pi \Phi_\mathrm{E}}{\Phi_0}\right)^4  + \mathcal{O}\left( \left(\frac{2 \pi \Phi_\mathrm{E}}{\Phi_0}\right)^6 \right) \right]
\end{split}
\end{equation}

\noindent Restricting the qubit operators to the first two levels results in the following Hamiltonian.

\begin{equation}
\label{eq:emitterMM Hamiltonian}
\begin{split}
\hat{H}/\hbar &= \frac{\omega_\mathrm{E}}{2}\hat{\sigma}^z_{\mathrm{E}} + \sum_{j=j_\mathrm{R}, j_\mathrm{L}} g \left(\hat{\sigma}^+_\mathrm{E}\hat{L}_j + \hat{\sigma}^-_\mathrm{E}\hat{L}_j^\dagger \right) \\
& \sum_j \omega_{0,j} \hat{L}_j^\dagger \hat{L}_j + J_{j,j+1} \left(\hat{L}_j^{\dagger} \hat{L}_{j+1} + \hat{L}_j \hat{L}_{j+1}^\dagger \right)
\end{split}
\end{equation}

\noindent Here, $\hat{\sigma}^{-(+)}_\mathrm{E}$ is the emitter qubit lowering (raising) operator, $\hat{\sigma}^z_\mathrm{E}$ is the emitter Pauli-Z operator, and $\hat{L}_j^{(+)}$ is the lowering (raising) operator for CCA mode $j$, which is either a coupler or a linear resonator. 

\begin{equation}
\hat{L}_j^{(+)} =
    \begin{cases}
        \hat{\sigma}^{-(+)}_{\mathrm{C},j} & j = j_\mathrm{L}, j_\mathrm{R} \\ 
        \hat{a}_j^{(\dagger)} & \mathrm{otherwise.}
    \end{cases}
\end{equation}

\noindent The CCA resonator frequencies remain unchanged at $\omega_{0,j} = 1/\sqrt{L_0C_\mathrm{tot}}$. Emitter and coupler frequencies follow standard transmon expressions. The emitter frequency is $\omega_\mathrm{E} = (\sqrt{8E_{J,\mathrm{E}} E_{C,\mathrm{E}}} - E_{C,\mathrm{E}})/{\hbar}$, and the coupler frequencies are $\omega_{0,j} = (\sqrt{8E_{J,\mathrm{C}} E_{C,j}} - E_{C,j})/{\hbar}$, where $j=j_\mathrm{L}$, $j_\mathrm{R}$. Here, $E_{C,\mathrm{E}} = e^2/2C_{\mathrm{tot},E}$ and $E_{C,j} = e^2/2C_{\mathrm{tot},j}$ are the charging energies of the emitter and coupler qubits. The emitter-coupler coupling rates are

\begin{equation}
g = \frac{C_g}{2\sqrt{C_{\mathrm{tot},E}C_{\mathrm{tot},j}}}\sqrt{\tilde{\omega}_\mathrm{E}\tilde{\omega}_{0,j}}
\end{equation}

\noindent for $j = j_\mathrm{L}$, $j_\mathrm{R}$, where we define the linearized qubit frequencies $\tilde{\omega}_\mathrm{E} = \sqrt{8E_{J,\mathrm{E}} E_{C,\mathrm{E}}}$ and $\tilde{\omega}_{0,j} = \sqrt{8E_{J,\mathrm{C}} E_{C,j}}$. The CCA hopping energy is 

\begin{equation}
J_{j,j+1} = \frac{C_{J,1}}{2\sqrt{C_{\mathrm{tot},j}C_{\mathrm{tot},j+1}}}\sqrt{\tilde{\omega}_{0,j}\tilde{\omega}_\mathrm{0,j+1}}.
\end{equation}

\noindent For linear CCA resonators, $\omega_{0,j} = \tilde{\omega}_{0,j}$.

We take emitter parameters to be $C_{E} = 83$ fF and $E_{J,E} = 8.34$ GHz, resulting in expected emitter frequency of $\omega_\mathrm{E}/2\pi = 3.35$ GHz. The couplers are impedance matched to the remainder of the CCA (CCA parameters: $L_0 = 6.5$ nH, $C_0 = 125$ fF, and $C_{J,1}= 24$ fF). We keep the coupler-coupler and coupler-resonator mutual capacitances uniform with the linear chain, at $C_{J,1}$. The coupler bare self-capacitance is $C_{0,C} = 115.0$ fF and the emitter-coupler capacitance is $C_{g} = 10.0$ fF. We choose $C_{0,C}+C_g = C_0$ to keep the non-hopping portion of the capacitances uniform between the couplers and resonators. Because we use flux-tunable coupler transmons, $E_{J,C}$ can be tuned in situ by a DC bias to align the frequencies of couplers to the resonators. We choose a maximum Josephson energy of $E_{J,C,\mathrm{max}} = 45$ GHz, which corresponds to a maximum coupler frequency of $\approx 6.2$ GHz. To tune up the CCA, we bias the coupler SQUID loops to $E_{J,C} \approx 25$ GHz, corresponding to an effective linear inductance of $L_{C}$ $\left(= {(\hbar/2e)^2}/{E_{J,C}}\right)\approx 6.5$ nH. 

To design the coupler circuit shown in \cref{fig:Fig1}d, we first perform an electromagnetics simulation of four CCA resonator unit cells with capacitive coupling to input-output ports. We previously modeled this chain using the ABCD matrix formalism, shown in \cref{fig:linearMM}g. We then replace the two central resonators with coupler elements, creating a resonator-coupler-coupler-resonator chain. Each coupler element consists of a metal island which is shunted in simulations by an ideal inductor placed at the target location for the coupler SQUID loop. By adjusting the geometry of the coupler's capacitor pads, we tune couplings to their desired values. This is shown in \cref{fig:embedded coupler}c, which depicts the simulated transmission of four coupled linear resonators in blue and the resonator-coupler-coupler-resonator chain in black. The two traces are nearly identical, indicating uniform coupling capacitances $C_{J,1}$ and non-hopping capacitances ($C_0$ for linear resonators, $C_{0,C}+C_g$ for couplers).

To distinguish between the coupler's bare self-capacitance $C_{0,C}$ and the emitter-coupler mutual capacitance $C_g$, we perform capacitive network simulations in which inductive components (ideal inductors and meandered wires) are removed. By driving each metal island with a separate co-calibrated input-output port, we can extract mutual capacitances directly between qubits using admittance ($Y$) parameters. In this case, two ports (1 and 2) driving separate metal islands form a $\pi$-network, and the mutual capacitance $C_{12} = -Y_{12}/i\omega$, where $Y_{12}$ is the admittance parameter between ports 1 and 2 and $\omega$ is the drive frequency.  

\begin{figure*}[t!]
\centering
\includegraphics[width=1\linewidth]{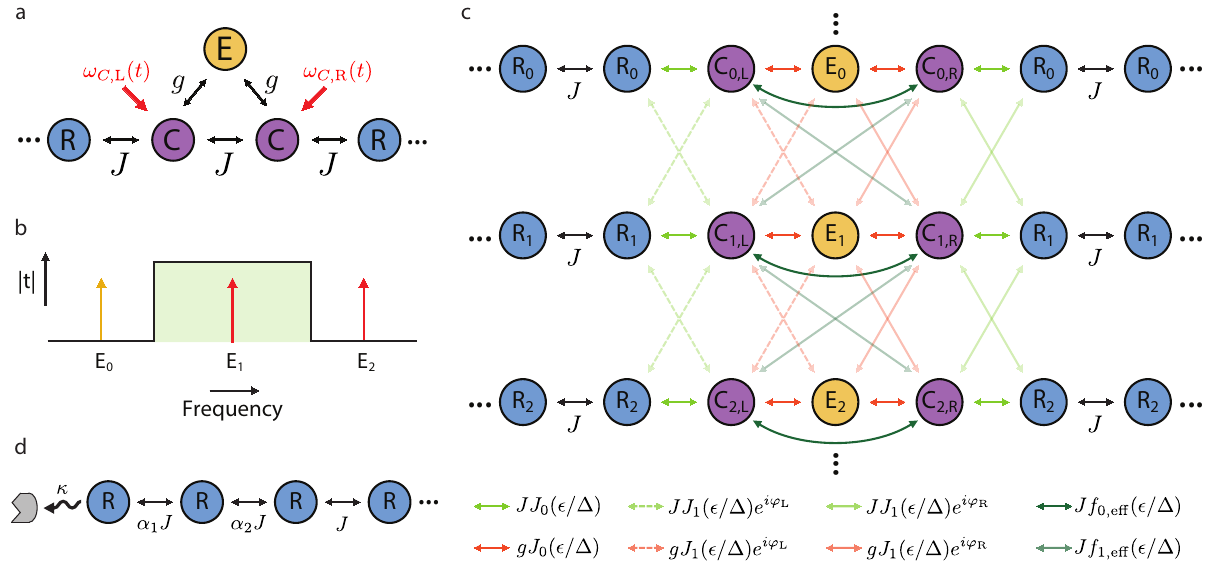}
\caption{\textbf{Parametric coupling model.} (a) Schematic of emitter coupled to coupled cavity array waveguide, with frequency modulation of the couplers (reproduced from \cref{fig:embedded coupler}a). (b) Diagram of CCA filtering showing the natural frequency of the emitter, $E_0$, outside of the passband. Activating the parametric coupling with a coupler modulation of frequency $\Delta$ creates emitter sidebands $E_1$ and $E_2$. $E_1$ falls in the passband and carries a parametric phase (either $\varphi_\mathrm{L}$ or $\varphi_\mathrm{R}$) while $E_2$ falls outside of the passband. (c) Frequency domain picture of emitter coupled to the CCA in the presence of parametric coupling. Arrows indicate effective interaction strengths between modes. The loop between $E_1$, $C_{0,\mathrm{L}}$, and $C_{0,\mathrm{R}}$ generates a chiral qubit-waveguide interaction. (d) Taper section of the CCA, with increased resonator-resonator coupling for modes near the edge of the chain. The first (last) resonator is coupled to an output port with rate $\kappa$.}
\label{fig:CMT}
\end{figure*}

\subsection{Parametric coupling model}
\label{appendix:Design:CMT}

Parametric coupling between the emitter and waveguide is generated by modulating the frequencies of coupler elements in the CCA, shown in \cref{fig:CMT}a. Harmonic modulation using a pump of frequency $\Delta$ creates sidebands of the emitter qubit at $\omega_\mathrm{E}+n\Delta$ ($n \in \mathbb{Z}$), shown in \cref{fig:CMT}b. The emitter's natural frequency ($\omega_\mathrm{E}$) sits outside of the passband of the CCA; we call this the emitter baseband, labeled $E_0$. The emitter's first sideband, $E_1$, lies inside the passband and can emit to the waveguide at frequency $\omega_\mathrm{E} + \Delta$. All other sidebands, like the second sideband ($E_2$, at frequency $\omega_\mathrm{E}+2\Delta$), lie outside of the passband and do not emit. Here, we extend a previous model for parametric coupling \cite{joshi2023resonance} to a chiral emitter coupled to a CCA. The generated sidebands and resulting effective interaction rates are shown in \cref{fig:CMT}c. The chiral emitter-waveguide interaction arises due to a synthetic gauge field in the loop containing $E_1$, $C_{0,\mathrm{L}}$, and $C_{0,\mathrm{R}}$, as discussed in the main text and \cref{appendix:Design:chiral}. We consider the Hamiltonian given in \cref{eq:emitterMM Hamiltonian} with $N$ CCA modes. For simplicity, we use linear bosonic operators for all modes. We also include a CCA taper section, as discussed in \cref{appendix:Design:circuit} and \cite{sumetsky2003modeling}. Setting $\hbar=1$, the Hamiltonian is given below.

\begin{equation}
    \hat{H} = \hat{H}_\mathrm{E} + \hat{H}_\mathrm{MM} + \hat{H}_\mathrm{bath} + \hat{H}_\mathrm{int}
\end{equation}

\begin{equation}
\hat{H}_\mathrm{E} = \omega_\mathrm{E} \hat{a}^\dagger_{\mathrm{E}} \hat{a}_\mathrm{E} + \sum_{j=j_\mathrm{R}, j_\mathrm{L}} g \left(\hat{a}^\dagger_\mathrm{E}\hat{a}_j + \hat{a}_\mathrm{E}\hat{a}_j^\dagger \right)
\end{equation}

\begin{equation}
\hat{H}_\mathrm{MM} = \sum^{N}_{j=1} \omega_{j}(t) \hat{a}_j^\dagger \hat{a}_j + \sum^{N-1}_{j=1} \alpha_j J \left(\hat{a}_j^{\dagger} \hat{a}_{j+1} + \hat{a}_j \hat{a}_{j+1}^\dagger \right)
\end{equation}

\begin{equation}
    \hat{H}_\mathrm{bath} = \sum_q\omega_q (\hat{b}^\dagger_{\mathrm{L},q}\hat{b}_{\mathrm{L},q} + \hat{b}^\dagger_{\mathrm{R},q}\hat{b}_{\mathrm{R},q})
\end{equation}

\begin{equation}
    \hat{H}_\mathrm{int} = \sum_q \left(-if_q \hat{b}_{\mathrm{L},q} \hat{a}^\dagger_1 -if_q \hat{b}_{\mathrm{R},q} \hat{a}_N^\dagger +h.c.\right)
\end{equation}

\noindent Here, $j_\mathrm{L}$ and $j_\mathrm{R}$ are the indices of the coupler modes, which we embed in the middle of the CCA, letting $j_\mathrm{L} = \lfloor \frac{N}{2} \rfloor$ and $j_\mathrm{R} = j_\mathrm{L}+1$. The coupling rates $g$ and $J$ are kept static, but the coupler frequencies are made time-dependent. The frequencies of CCA modes are set to

\begin{equation}
\omega_j(t) = 
    \begin{cases}
        \omega_0 +\epsilon \sin(\Delta t +\varphi_\mathrm{L}) & j=j_\mathrm{L} \\
        \omega_0 +\epsilon \sin(\Delta t +\varphi_\mathrm{R}) & j=j_\mathrm{R} \\
       \omega_0 & j \neq j_\mathrm{L}, j_\mathrm{R}. 
    \end{cases}
\end{equation}

\noindent To create a tapering section between the CCA and the two input-output modes at the opposite ends of the CCA, we vary the static hopping rates between modes near the edges of the resonator chain as shown in \cref{fig:CMT}d; this is denoted by $\alpha_j$. We set $\alpha_j = 1.35$ for $j=1,N-1$; $\alpha_j = 1.04$ for $j=2,N-2$; and $\alpha_j=1$ otherwise \cite{sumetsky2003modeling}. The modes of the left and right waveguide baths are given in $\hat{H}_\mathrm{bath}$ and couple to the first and last CCA resonators as described in $\hat{H}_\mathrm{int}$. The resonator-bath interaction produces a decay rate of $\kappa$ to the bath for the first and last resonators, as shown in \cref{fig:CMT}d. We then write the Langevin equations for the emitter and all CCA modes, followed by performing the unitary transformation $\hat{\tilde{a}}_{j_\mathrm{L(R)}} = U_\mathrm{L(R)}^\dagger \hat{a}_{j_\mathrm{L(R)}}U_\mathrm{L(R)}$, where

\begin{equation}
    U_\mathrm{L(R)} = \exp \left[i\frac{\epsilon}{\Delta} \cos(\Delta t + \varphi_\mathrm{L(R)}) \hat{a}_{j_\mathrm{L(R)}}^\dagger\hat{a}_{j_\mathrm{L(R)}}\right]
\end{equation}

\begin{widetext}

\noindent This is equivalent to performing the following substitution.

\begin{equation}
\hat{\tilde{a}}_{j_\mathrm{L (R)}} = \hat{a}_{j_\mathrm{L(R)}} \exp \left[-i\frac{\epsilon}{\Delta}\cos(\Delta t + \varphi_\mathrm{L(R)}) \right]
\end{equation}

\noindent These unitary transformations allow us to then expand the time dependence of the modified coupler operator into discrete harmonics using the Jacobi-Anger expansion, where $J_n$ is the $n$'th Bessel function of the first kind.

\begin{equation}
    \exp \left[ i \frac{\epsilon}{\Delta} \cos(\Delta t +\varphi_\mathrm{L(R)}) \right] = \sum_{n=-\infty}^\infty i^n J_n(\frac{\epsilon}{\Delta})e^{in(\Delta t +\varphi_\mathrm{L(R)})}
\end{equation}

\noindent The equation of motion for each mode can then be Fourier transformed, and each mode can be separated into discrete sidebands with frequency spacing $\Delta$. We can then write an equation of motion for each sideband of each mode, treating sidebands independently. Each sideband order $n$ (existing at frequency $\omega_j+n\Delta$) then contains $N+1$ equations of motion (describing the $n$'th sideband of $N$ CCA modes and one emitter), resulting in the coupled-mode matrix equation given below.

\begin{equation}
\label{eq:CMT}
    -i\sqrt{\kappa}
    \begin{bmatrix}
        \vdots \\
        \mathbf{0} \\
        \mathbf{\hat{a}_\mathrm{in}} \\
        \mathbf{0} \\
        \vdots \
    \end{bmatrix} 
    = 
    \begin{bmatrix}
        \ddots & \vdots & \vdots & \vdots & \iddots \\
        \dots & \mathbf{H_{-1}} & \mathbf{G_{1}} & \mathbf{G_{2}} & \dots \\
        \dots & \mathbf{G_{1}^*} & \mathbf{H_{0}} & \mathbf{G_{1}} & \dots \\
        \dots & \mathbf{G^*_{2}} & \mathbf{G_{1}^*} & \mathbf{H_{1}} & \dots \\ 
        \iddots & \vdots & \vdots & \vdots & \ddots \\
    \end{bmatrix}
    \begin{bmatrix}
        \vdots \\
        \mathbf{\hat{a}(\omega-\Delta)} \\
        \mathbf{\hat{a}(\omega)} \\
        \mathbf{\hat{a}(\omega+\Delta)} \\
        \vdots \\
    \end{bmatrix}
\end{equation}

\noindent Here, $\mathbf{\hat{a}_\mathrm{in}}$ is a $(N+1)\times1$ column vector corresponding to the input modes at the edges of the CCA and is defined as $\mathbf{\hat{a}_\mathrm{in}}= [\hat{a}_\mathrm{in,L},0,...,0,\hat{a}_\mathrm{in,R}]^\intercal$. $\mathbf{0}$ is a $(N+1)\times1$ zero vector. $\mathbf{\hat{a}(\omega+n\Delta)}$ is a $(N+1)\times1$ column vector containing the $N+1$ modes of sideband order $n$. The vector is defined as $\mathbf{\hat{a}(\omega+n\Delta)} = [\hat{a}_1(\omega+n\Delta), ... , \hat{\tilde{a}}_{j_\mathrm{L}}(\omega+n\Delta), \hat{a}_\mathrm{E}(\omega+n\Delta), \hat{\tilde{a}}_{j_\mathrm{R}}(\omega+n\Delta), ... , \hat{a}_{N}(\omega+n\Delta), ]^\intercal$; the entries contain the CCA modes in increasing index order, with the emitter inserted between the coupler modes. The sub-matrices of the effective Hamiltonian matrix $\mathbf{H_\mathrm{eff}}$ are $\mathbf{H}_n$ and $\mathbf{G}_m$. $\mathbf{H}_n$ describes the on-site frequency and decay rates of each mode of sideband order $n$, as well as interactions within the sideband order. Meanwhile, $\mathbf{G}_m$ describes the interaction rates between modes of different sideband orders. Here, $m = |n - n'|$ is the separation between two sideband orders. For example, $\mathbf{H_0}$ and $\mathbf{H_1}$ are separated by $m=1$ and are therefore coupled together by sub-matrix $\mathbf{G}_1$. $\mathbf{G}^*_m$ is the conjugate transpose of $\mathbf{G}_m$. 

We comment that interaction terms in the coupled-mode sideband picture are scaled by the Bessel functions $J_n(\epsilon/\Delta)$. For small $\epsilon/\Delta < 1$ , higher order Bessel functions take on vanishingly small values. Additionally, interaction terms contained in a coupling sub-matrix $\mathbf{G}_m$ generally contains terms scaled by $J_m(\epsilon/\Delta)$. These two facts allow us to truncate the infinite matrix equation given in \cref{eq:CMT}. For simplicity, we will ignore interactions containing $J_m(\epsilon/\Delta)$ where $m\geq2$. We will also ignore interactions that are second order in $J_m(\epsilon/\Delta)$ where $m\geq1$. Lastly, we will also only consider sideband orders for $n \in\{-2,-1,0,1,2\}$. We can then explicitly write the sub-matrices $\mathbf{H_n}$ and $\mathbf{G_m}$.

\begin{equation}
\mathbf{H_n} = 
\begin{bmatrix}
\delta_{1,n}+i\frac{\kappa +\gamma_1}{2} & -\alpha_1 J  & 0 & 0 & \dots & \dots & \dots & \dots & 0 \\

-\alpha_1 J & \ddots & \ddots & \ddots & \ddots & & & & \vdots \\

0 & \ddots & \ddots &-J J_0(\frac{\epsilon}{\Delta}) & 0 & \ddots & & & \vdots\\

0 & \ddots & -J J_0(\frac{\epsilon}{\Delta}) & \delta_{j_\mathrm{L}}+i\frac{\gamma_{j_\mathrm{L}}}{2} & -gJ_0(\frac{\epsilon}{\Delta}) & -Jf_{0,\mathrm{eff}}(\frac{\epsilon}{\Delta}) & \ddots & & \vdots\\

\vdots & \ddots & 0 & -gJ_0(\frac{\epsilon}{\Delta}) & \delta_\mathrm{E} + i\frac{\gamma_\mathrm{E}}{2} & -g J_0(\frac{\epsilon}{\Delta}) & 0 & \ddots & \vdots \\

 \vdots & & \ddots & -Jf_{0,\mathrm{eff}}\left(\frac{\epsilon}{\Delta}\right) & -g J_0(\frac{\epsilon}{\Delta}) & \delta_{j_\mathrm{R}}+i\frac{\gamma_{j_\mathrm{R}}}{2} & -J J_0(\frac{\epsilon}{\Delta}) & \ddots & 0 \\

 \vdots & & & \ddots & 0 & -J J_0(\frac{\epsilon}{\Delta}) & \ddots& \ddots & 0 \\

\vdots & & & & \ddots & \ddots & \ddots & \ddots & -\alpha_{N-1} J\\

0 & \dots & \dots & \dots& \dots & 0 & 0 & -\alpha_{N-1} J& \delta_{N,n}+i\frac{\kappa +\gamma_N}{2}\\
\end{bmatrix}
\end{equation}

\noindent Here, $\delta_{j,n} = \omega - (\omega_j + n\Delta)$ is the detuning of the $n$'th sideband of mode $j$ and $\gamma_j$ is the internal dissipation rate of mode $j$. The first and $N$'th CCA mode are coupled to external ports with rate $\kappa$. We comment that the interaction rates between modes of $\mathbf{H_n}$ have the same structure as the static Hamiltonian in the absence of frequency modulation. However, the interactions of each coupler are diluted by the zeroth order Bessel function $J_0(\epsilon/\Delta)$. The coupler-coupler interaction is further diluted, for which we define a zeroth order dilution factor $f_{0,\mathrm{eff}}(\epsilon/\Delta)$, defined below. 

\begin{equation}
    f_{0,\mathrm{eff}} \left(\frac{\epsilon}{\Delta} \right) = \left[J_0\left(\frac{\epsilon}{\Delta}\right) \right]^2 + \sum_{n=1}^\infty 2 \left[ J_n\left(\frac{\epsilon}{\Delta}\right)\right]^2\cos(n\Delta\varphi)
\end{equation}

\noindent For small $\epsilon/\Delta$, we can discard terms of order $\left[ J_1(\epsilon/\Delta) \right]^2$ and above. The coupling terms defined in $\mathbf{H_n}$ are depicted in \cref{fig:CMT}c as interactions within each row, as each row represents a sideband order. We also define the coupling matrix $\mathbf{G_1}$ below, where all non-zero elements are listed explicitly.

\begin{equation}
    \mathbf{G_1} = 
    \begin{bmatrix}
        0 & 0 & 0 & 0 & \dots & \dots & \dots & \dots & \dots & \dots & 0\\
        0 & \ddots & \ddots & \ddots & \ddots &  & & & & & \vdots \\ 
        0 & \ddots & \ddots & 0 & \ddots & \ddots & & & & & \vdots \\ 
        0 & \ddots & 0& 0 & -iJJ_1(\frac{\epsilon}{\Delta}) e^{-i\varphi_\mathrm{L}} & 0 & \ddots & & & & \vdots\\
        \vdots & \ddots & \ddots & iJJ_1(\frac{\epsilon}{\Delta}) e^{-i\varphi_\mathrm{L}} & 0  & igJ_1(\frac{\epsilon}{\Delta}) e^{-i\varphi_\mathrm{L}} & -iJ f_{1,\mathrm{eff}}(\frac{\epsilon}{\Delta}) & \ddots & & & \vdots \\
        \vdots & & \ddots & 0 & -igJ_1(\frac{\epsilon}{\Delta}) e^{-i\varphi_\mathrm{L}} & 0 & -igJ_1(\frac{\epsilon}{\Delta}) e^{-i\varphi_\mathrm{R}} & 0 & \ddots & & \vdots\\
        \vdots & & & \ddots & iJ f_{1,\mathrm{eff}}(\frac{\epsilon}{\Delta})&igJ_1(\frac{\epsilon}{\Delta}) e^{-i\varphi_\mathrm{R}} & 0 & iJJ_1(\frac{\epsilon}{\Delta}) e^{-i\varphi_\mathrm{R}} & \ddots & \ddots & \vdots \\
        \vdots & & & & \ddots & 0 & -iJJ_1(\frac{\epsilon}{\Delta}) e^{-i\varphi_\mathrm{R}} & 0 & 0 & \ddots & 0 \\
        \vdots & & & & & \ddots & \ddots & 0 & \ddots & \ddots & 0 \\
        \vdots & & & & & & \ddots & \ddots & \ddots & \ddots & 0 \\
        0 & \dots & \dots & \dots &\dots & \dots & \dots & 0 & 0 & 0 & 0 \\
    \end{bmatrix}
\end{equation}

\noindent We observe that coupling terms are diluted by $J_1(\epsilon/\Delta)$, with the coupler-coupler interaction diluted by the first order dilution factor $f_{1,\mathrm{eff}}(\epsilon/\Delta)$. We define $f_{1,\mathrm{eff}}(\epsilon/\Delta)$ to first order below, discarding terms containing $J_m(\epsilon/\Delta)$ for $m\geq2$.

\begin{equation}
    f_{1,\mathrm{eff}}\left(\frac{\epsilon}{\Delta}\right) \approx J_0\left(\frac{\epsilon}{\Delta}\right) J_1\left(\frac{\epsilon}{\Delta}\right) \left(e^{-i\varphi_\mathrm{R}}-e^{-i\varphi_\mathrm{L}}\right)
\end{equation}

\noindent The coupling elements depicted in $\mathbf{G_1}$ are depicted in \cref{fig:CMT}c as interactions between rows, shown using diagonal arrows. $\mathbf{G_m}$ is a $(N+1)\times(N+1)$ zero matrix for $m\geq2$ in our small $\epsilon/\Delta$ approximation. Pictorially, including these higher order interactions would add additional diagonal arrows crossing multiple rows in \cref{fig:CMT}c.

The equations of motion described above can be combined with input-output relations to compute the full device transmission. The input-output relation is given as 

\begin{equation}
    \hat{a}_\mathrm{out,R} = -\sqrt\kappa \hat{a}_N(\omega)
\end{equation}

\noindent We can solve for $\hat{a}_N(\omega)$ by inverting the effective Hamiltonian $\mathbf{H_\mathrm{eff}}$. We provide an example of the transmission of a single chiral qubit coupled to a waveguide using the coupled-mode model in \cref{fig:CMT transmission}. \cref{fig:CMT transmission}a shows the CCA passband with modulation turned on for a forward and backward chiral qubit. \cref{fig:CMT transmission}b (c) shows a zoomed-in trace of the transmission (phase) of the chiral qubit in the forward and backward settings, with fits to \cref{eq:transmission} applied. 

Using the coupled-mode model, we find that $\epsilon/\Delta \approx 0.31$ for our experimental device. For a forward chiral phase setting and representative settings of $\omega_\mathrm{E} = 3.4$ GHz, $g = 167$ MHz, and $\Delta = 1.3$ GHz, we find that this drive amplitude corresponds to $\Gamma_\mathrm{1D}^f = 2$ MHz. For this drive setting, $J_0(\epsilon/\Delta) = 0.98$, $J_1(\epsilon/\Delta) = 0.15$, and $J_2(\epsilon/\Delta) = 0.01$, justifying the truncation of the matrix equation \cref{eq:CMT}. Lastly, we comment that the coupling between the first emitter sideband $E_1$ and either of the two couplers, $C_{0,\mathrm{L}}$, and $C_{0,\mathrm{R}}$, scales as $\sim gJ_1(\epsilon/{\Delta})$, which we replace with $g_\mathrm{eff}$ in the main text and \cref{fig:Fig1}b.

\end{widetext}

\begin{figure}
\centering
\includegraphics[width=1\linewidth]{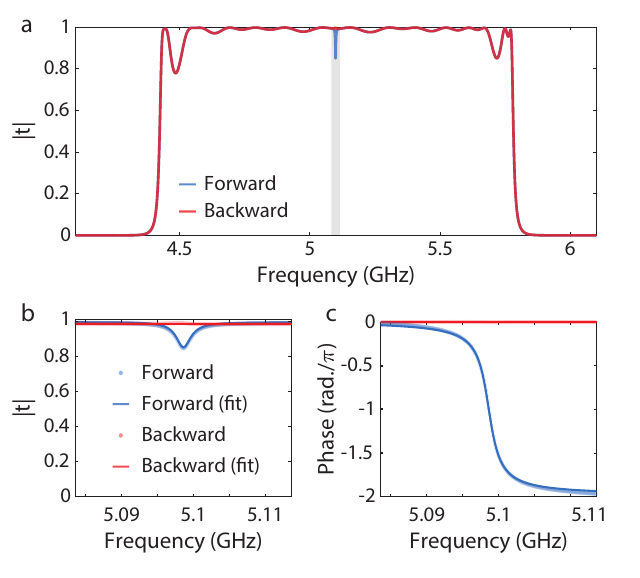}
\caption{\textbf{Transmission calculated using parametric coupling model.} (a) Transmission for an emitter coupled to a coupled cavity array with $N = 16$ modes and $J=0.35$ GHz. The parameters used are $\omega_\mathrm{E} = 3.6$ GHz, $g=167$ MHz, $\Delta = 1.53$ GHz, and $\epsilon/\Delta = 0.3$.  We set the emitter internal dissipation rate $\gamma_\mathrm{E} = 200$ kHz. The blue (red) trace shows a forward (backward) chiral emitter. (b) Zoomed-in trace of transmission and (c) phase for the forward (backward) chiral emitter shown in blue (red), with fits to \cref{eq:transmission}.}
\label{fig:CMT transmission}
\end{figure}

\section{Modeling}
\label{appendix:Modeling}

\subsection{Low-power cascaded qubit transmission using ABCD matrices}
\label{appendix:Modeling:ABCD}

When probed at sufficiently low powers ($\Omega \ll \Gamma_\mathrm{1D}^f$), the qubits are negligibly populated and do not power broaden. In this regime, the qubits can be effectively modeled as a chain of linear chiral cavities. Provided that all of the qubits are perfectly chiral ($\Gamma^f_\mathrm{1D} \neq0$, $\Gamma^b_\mathrm{1D}= 0$), the transmission of a k-qubit cascaded chain is then simply the product of the each individual qubit's transmission, $t_\mathrm{tot} = t_1 * ... *t_{N-1}*t_N$. This is readily apparent if we consider that there is no information backflow or inelastic scattering in the system. This simple transmission relation no longer holds true in the case of imperfect chirality ($\Gamma^b_\mathrm{1D} \neq 0$) because emission in the backward direction can cause reflections in the chain. The backward decay rates ($\Gamma_\mathrm{1D}^b$), waveguide phase between qubits, and loss between qubits must then all be accounted for. For the low-power, linear regime, this can be done using the ABCD matrix (transmission matrix) formalism discussed in \cref{appendix:Design:circuit}. To represent $N$ cascaded chiral qubits with imperfect chirality, accounting for waveguide phase and loss between the qubits, we can then use the following product of ABCD matrices

\begin{equation}
\mathrm{M}_\mathrm{tot} = \mathrm{M}_\mathrm{1}*\mathrm{M}_\mathrm{1,2}*\mathrm{M}_\mathrm{2}* ... *\mathrm{M}_{N-1,N}*\mathrm{M}_{N}
\label{eq:ABCD_system}
\end{equation}

\noindent where $\mathrm{M}_\mathrm{tot}$ is the ABCD matrix for the entire qubit chain, $\mathrm{M}_{j}$ is the ABCD matrix for qubit $j$, and $\mathrm{M}_{j,k}$ is the ABCD matrix for the waveguide section between qubits $j$ and $k$. $\mathrm{M}_{j,k}$ is defined as  

\begin{equation}
\begin{split}
\mathrm{M}_{j,k} & = \begin{bmatrix} A_{j,k} &B_{j,k} \\ C_{j,k} & D_{j,k} \end{bmatrix}\\
& = \begin{bmatrix} \cos(\tilde{\varphi}_{j,k}) & iZ_0\sin(\tilde{\varphi}_{j,k}) \\ i\sin(\tilde{\varphi}_{j,k})/Z_0 & \cos(\tilde{\varphi}_{j,k})  \end{bmatrix} 
\label{eq:ABCD_cascade}
\end{split}
\end{equation}

\noindent where $Z_0$ is the characteristic impedance of the waveguide and $\tilde{\varphi}_{j,k} = \varphi_{j,k}+ i\log(\eta_{j,k})$. Here, $\varphi_{j,k}$ is the phase accumulated by a photon between qubits $j$ and $k$, and $\eta_{j,k}^2$ is the probability of a photon at qubit $j$ reaching qubit $k$. To obtain the ABCD matrix for a single qubit, we make use of the mapping between ABCD and S-matrices \cite{pozar2011microwave}. We use the following S-matrix

\begin{equation}
\begin{split}
\mathrm{S}_{j} & = \begin{bmatrix} S_{11,j} & S_{12,j} \\ S_{21,j} & S_{22,j}\end{bmatrix} \\
& = \begin{bmatrix} 
\frac{\sqrt{\Gamma^f_\mathrm{1D} \Gamma_\mathrm{1D}^b}}{i\Delta+\Gamma_\mathrm{tot}/2} & 
1-\frac{\Gamma_\mathrm{1D}^b}{i\Delta+\Gamma_\mathrm{tot}/2} \\ 
1-\frac{\Gamma_\mathrm{1D}^f}{i\Delta+\Gamma_\mathrm{tot}/2} & 
\frac{\sqrt{\Gamma^f_\mathrm{1D} \Gamma_\mathrm{1D}^b}}{i\Delta+\Gamma_\mathrm{tot}/2}
\end{bmatrix} 
\label{eq:Smatrix}
\end{split}
\end{equation}

\noindent where S-parameters are derived using input-output theory \cite{joshi2023resonance}. By mapping $\mathrm{M}_\mathrm{tot}$ to an S-matrix, we obtain the transmission coefficient $S_{21,\mathrm{tot}}$ of the qubit chain. Because $S_{21,\mathrm{tot}}$ is over-parameterized in the case of many qubits, we apply independent fits to the transmission of individual qubits and use the extracted $\Gamma_\mathrm{tot}$ and $\Gamma_\mathrm{1D}^f$ as inputs to fits of larger chains. We then simultaneously fit the transmission of two-, three-, and four-qubit chains to the derived $S_{21,\mathrm{tot}}$ to obtain the plots shown in \cref{fig:Fig1}e,f,g,h.

\subsection{SLH formalism and master equation}
\label{appendix:Modeling:SLH ME}

In the general case, the ABCD matrix formalism is not sufficient to capture non-linearity in the cascaded qubit chain. To properly model the qubit chain, we use the SLH formalism \cite{combes2017slh} to derive the master equation and input-output relations. The SLH formalism allows for modeling open quantum systems by combining their constituent sub-systems, where each sub-system is represented by an SLH triplet $G = (\mathbf{S},\mathbf{L},H)$. For a system with $n$ input-output ports, $\mathbf{S}$ is an $n\times n$ scattering matrix, $\mathbf{L}$ is an $n \times 1$ vector representing the coupling to each port, and $H$ is the Hamiltonian. To model the four-qubit chain (including imperfect chirality, waveguide loss, and waveguide phase), the SLH triplet for qubit $j$ in the drive frame is 

\begin{equation}
G_{\mathrm{Q},j}^{f (b)} = \left( \mathbbm{1}_4, \begin{bmatrix} \sqrt{\Gamma^{f(b)}_{\mathrm{1D},j}}\hat{\sigma}^-_{j} \\ 0 \\ 0 \\ 0 \end{bmatrix}, -\frac{\delta_j}{2}\hat{\sigma}_{j}^z  \right)
\end{equation}

\noindent for forward (backward) propagation in the waveguide. For a given direction of propagation, there are 4 input-output ports: the output of the waveguide, and three output ports that account for photon loss during propagation between individual qubits. The scattering matrix $\mathbbm{1}_4$ is therefore a $4\times4$ identity matrix. $\Gamma^{f(b)}_{\mathrm{1D},j}$ is qubit $j$'s waveguide decay rate in the forward (backward) direction, $\delta_j = \omega_\mathrm{D}-\omega_{\mathrm{Q},j}$ is the detuning between the drive ($\omega_\mathrm{D}$) and qubit ($\omega_{\mathrm{Q},j}$) frequencies, $\hat{\sigma}_j^-$ is the qubit lowering operator, and $\hat{\sigma}_{j}^z$ is the qubit Pauli-Z operator. The SLH triplet for the section of waveguide between qubits $j$ and $k$ is

\begin{equation}
G_{j,k}^{f(b)} = \left( \mathbf{S}_{\bm{j,k}}, \mathbf{0}, 0 \right)
\label{eq:SLH_wg12}
\end{equation}

\noindent where $\mathbf{S}_{\bm{j,k}}$ is a $4\times4$ scattering matrix accounting for waveguide loss via a beam-splitter \cite{Irfan2024Mar}, and $\mathbf{0}$ is a $4\times 1$ zero vector. The scattering matrices for each waveguide section are

\begin{equation}
\mathbf{S_{1,2}} = \begin{bmatrix} 
\eta_{1,2} e^{i\varphi_{1,2}} & 0 & 0 & e^{i\varphi_{1,2}}\sqrt{1-\eta_{1,2}^2} \\ 
0 & 1 & 0 & 0 \\ 
0 & 0 & 1 & 0 \\ 
e^{i\varphi_{1,2}}\sqrt{1-\eta_{1,2}^2} & 0& 0 & \eta_{1,2}e^{i\varphi_{1,2}} \\ 
\end{bmatrix}
\end{equation}

\begin{equation}
\mathbf{S_{2,3}} = \begin{bmatrix} 
\eta_{2,3} e^{i\varphi_{2,3}}& 0 & e^{i\varphi_{2,3}}\sqrt{1-\eta_{2,3}^2} & 0 \\ 
0 & 1 & 0 & 0 \\ 
e^{i\varphi_{2,3}}\sqrt{1-\eta_{2,3}^2}  & 0 & \eta_{2,3} e^{i\varphi_{2,3}} & 0 \\ 
0 & 0& 0 & 1 \\ 
\end{bmatrix}
\end{equation}

\begin{align}
\mathbf{S_{3,4}} = \begin{bmatrix} 
\eta_{3,4} e^{i\varphi_{3,4}}& e^{i\varphi_{3,4}}\sqrt{1-\eta_{3,4}^2} & 0 & 0 \\ 
e^{i\varphi_{3,4}}\sqrt{1-\eta_{3,4}^2} & \eta_{3,4} e^{i\varphi_{3,4}} & 0 & 0 \\ 
0  & 0 & 1 & 0 \\ 
0 & 0& 0 & 1 \\ 
\end{bmatrix}
\end{align}

\noindent where $\eta_{j,k}^2$ is the probability of a photon propagating from qubit $j$ to qubit $k$ and $\varphi_{j,k}$ is the phase accumulated between qubits $j$ and $k$. We also include a classical drive at the input of the waveguide in the forward direction with amplitude $\alpha$ and phase $\varphi_\mathrm{in}$. The SLH triplet for the input drive is given below.
\begin{equation}
G_\mathrm{in} = \left( \mathbbm{1}_4, \begin{bmatrix} \alpha e^{i\varphi_\mathrm{in}}  \\ 0 \\ 0 \\ 0 \end{bmatrix}, 0 \right)
\end{equation}

\noindent The total SLH triplets for the forward and backward propagating modes can then be obtained using the series product operation, as shown below.
\begin{equation}
G^f_\mathrm{tot} = G_{\mathrm{Q}4}^f \triangleleft G_{3,4}^f \triangleleft G_{\mathrm{Q}3}^f \triangleleft G_{2,3}^f\triangleleft G_{\mathrm{Q}2}^f \triangleleft G_{1,2}^f \triangleleft  G_{\mathrm{Q}1}^f \triangleleft G_\mathrm{in}
\end{equation}
\begin{equation}
G^b_\mathrm{tot} = G_{\mathrm{Q}1}^b \triangleleft G_{1,2}^b \triangleleft G_{\mathrm{Q}2}^b \triangleleft G_{2,3}^b\triangleleft G_{\mathrm{Q}3}^b \triangleleft G_{3,4}^b \triangleleft  G_{\mathrm{Q}4}^b
\end{equation}
\noindent and the full system can be modeled using the concatenation product $G_\mathrm{tot} = G^f_\mathrm{tot} \boxplus G^b_\mathrm{tot}$. This allows us to derive the full master equation of the four-qubit chain, as given below.

\begin{equation}
\dot{\rho} = \dfrac{i}{\hbar}[\hat{H}_\mathrm{0} + \hat{H}^f_{\mathrm{int}}+H^b_\mathrm{int}, \rho] + \mathcal{L}\rho,
\end{equation}

\noindent The qubit and drive Hamiltonian is given by

\begin{widetext}

\begin{equation}
\hat{H}_\mathrm{0} = \sum_{j=1}^N -\frac{\delta_j}{2}\hat{\sigma}^z_{j} + \frac{i\Omega_{j} e^{-i\varphi_\mathrm{in}}}{2} \left(\prod_{k=0}^{j-1} \eta_{k,k+1}e^{-i\varphi_{k,k+1}} \right)\hat{\sigma}^-_j + h.c.
\end{equation}

\noindent where $N = 4$ is the number of qubits in the chain. $\Omega_{j} = 2 \alpha \sqrt{\Gamma_{{\mathrm{1D}},j}^f}$ is the Rabi drive frequency on qubit $j$. Assuming no inter-qubit waveguide loss or phase ($\eta_{k,k+1}e^{-{i\varphi_{k,k+1}}}=1$, for all $k$) and uniform qubit driving ($\Omega_j=\Omega$, for all $j$) without any drive phase ($\varphi_\mathrm{in}=0$), the Hamiltonian simplifies to 

\begin{equation}
\hat{H}_{\mathrm{0}} = \sum^{N}_{j=\mathrm{1}} -\frac{\delta_{j}}{2}\hat{\sigma}^z_{j} + \frac{i\Omega}{2}\left(\hat{\sigma}^{-}_{j} - \hat{\sigma}^+_{j} \right).
\label{eq:ideal-qubit-drive-hamiltonian}
\end{equation}

\noindent The interaction Hamiltonians are given by

\begin{equation}
\hat{H}_\mathrm{int}^f = \sum_{1 \leq j < k \leq N} \frac{i\sqrt{\Gamma^f_{\mathrm{1D},j}\Gamma^f_{\mathrm{1D},k}}}{2}\left(\prod_{j}^{k-1} \eta_{j,j+1}e^{-i\varphi_{j,j+1}} \right)\hat{\sigma}_j^{+}\hat{\sigma}^-_k + h.c.
\end{equation}

\begin{equation}
\hat{H}_\mathrm{int}^b = \sum_{1 \leq j < k \leq N} \frac{i\sqrt{\Gamma^b_{\mathrm{1D},j}\Gamma^b_{\mathrm{1D},k}}}{2}\left(\prod_{j}^{k-1} \eta_{j,j+1}e^{-i\varphi_{j,j+1}} \right)\hat{\sigma}^-_j \hat{\sigma}_k^{+} + h.c.
\end{equation}

\noindent where $\eta_{0,1} = 1$ and $\varphi_{0,1} = 0$ are dummy variables.  $H_\mathrm{int}^f$ and $H_\mathrm{int}^b$ represent photon-mediated interactions between pairs of qubits due to forward and backward qubit waveguide decay, respectively. If the qubits are fully bidirectional ($\Gamma_{\mathrm{1D},j}^f = \Gamma_{\mathrm{1D},j}^b$),  $H_\mathrm{int}^f$ and $H_\mathrm{int}^b$ simplify into the standard photon-mediated interactions of waveguide QED \cite{lalumiere2013waveguide}. If all qubits are perfectly chiral, $H_\mathrm{int}^b = 0$. Then, assuming no inter-qubit waveguide loss or phase ($\eta_{j,j+1}e^{-{i\varphi_{j,j+1}}}=1$, for all $j$) and uniform waveguide decay rates ($\Gamma^f_{\mathrm{1D},j} = \Gamma^f_\mathrm{1D}$, for all $j$), $\hat{H}^f_\mathrm{int}$ simplifies to

\begin{equation}
\hat{H}^f_{\mathrm{int}} = \sum_{1 \leq j < k \leq N} \frac{i\Gamma^f_\mathrm{1D}}{2} (\hat{\sigma}^{+}_{j}\hat{\sigma}^-_{k} - \hat{\sigma}^-_{j}\hat{\sigma}^{+}_{k}).
\label{eq:ideal-int-hamiltonian}
\end{equation}

\noindent The Liouvillian is given by 

\begin{equation}
\mathcal{L}\rho = \sum_{j=1}^N \mathcal{D}[\hat{F}_j]\rho +\sum_{j=1}^N \mathcal{D}[\hat{B}_j]\rho + \sum_{j=1}^N \Gamma'_{0,j}\mathcal{D}[\hat{\sigma}_j]\rho + \sum_{j=1}^N \frac{\Gamma_{\varphi,j}}{2}\mathcal{D}[\hat{\sigma}^{z}_{j}]\rho 
\end{equation}

\noindent where $\hat{F_j}$ and $\hat{B}_j$ are collapse operators corresponding to output ports for the forward and backward propagation in the waveguide. These include the loss channels included between each qubit and the waveguide output port. $\Gamma_{0,j}'$ is the internal decay rate of qubit $j$, and $\Gamma_{\varphi,j}$ is the pure dephasing of qubit $j$; these terms are added to the Lindbladian independent of the expression derived using the SLH formalism. The dissipator is given by $\mathcal{D}[A]\rho = A\rho A^\dagger - \dfrac{1}{2} A^\dagger A \rho - \dfrac{1}{2}\rho A^\dagger A$. The explicit forms of the waveguide collapse operators are given as

\begin{equation}
\hat{F}_j =
\begin{cases}
 e^{i\varphi_{j,j+1}} \sqrt{1 - \eta^2_{j,j+1}} 
\left[ \sqrt{\Gamma^f_{\mathrm{1D},j}}\hat{\sigma}^-_j +
\sum_{k=1}^{j-1} 
\left(\prod_{l=k}^{j-1} \eta_{l,l+1}e^{i\varphi_{l,l+1}} \right)
\sqrt{\Gamma^f_{\mathrm{1D},k}}\hat{\sigma}^-_k
\right] & j = 1,2,...,N-1 \\
\sum_{k=1}^j (\prod^{j-1}_{l=k} \eta_{l,l+1}e^{i\varphi_{l,l+1}} )\sqrt{\Gamma^f_{\mathrm{1D},k}}\hat{\sigma}^-_k & j=N
\end{cases}
\end{equation}

\begin{equation}
\hat{B}_j =
\begin{cases}
 e^{i\varphi_{N-j,N-j+1}} \sqrt{1 - \eta^2_{N-j,N-j+1}} \left[ \sqrt{\Gamma^f_{\mathrm{1D},N-j+1}}\hat{\sigma}^-_{N-j+1} \right. \\
\left. +\sum_{k=N-j+1}^{N-1} \left(\prod_{l=N-j+1}^{k} \eta_{l,l+1}e^{i\varphi_{l,l+1}} \right) \sqrt{\Gamma^f_{\mathrm{1D},k+1}}\hat{\sigma}^-_{k+1} \right] & j = 1,2,...,N-1 \\
\sum_{k=1}^j (\prod^{k-1}_{l=1} \eta_{l,l+1}e^{i\varphi_{l,l+1}} )\sqrt{\Gamma^b_{\mathrm{1D},k}}\hat{\sigma}^-_k & j=N \\
\end{cases}
\end{equation}

\end{widetext}

\noindent where $j=1, 2, ... , N-1$ corresponds to loss during propagation between qubits and $j=N$ corresponds to the output of the waveguide. To recover the Hamiltonians (\cref{eq:ideal-qubit-drive-hamiltonian} and \cref{eq:ideal-int-hamiltonian}) and Liouvillian given in the main text, we assume perfect chirality ($\Gamma_{\mathrm{1D},j}^b=0$ for $j\in{\{1,...,N-1,N\}}$), set the qubit waveguide decay rates to be equal $\Gamma_{\mathrm{1D},1}^f = \Gamma_{\mathrm{1D},2}^f = ... =  \Gamma_{\mathrm{1D},N}^f = \Gamma_{\mathrm{1D}}^f$, and assume no loss or phase accumulation between qubits ($\eta_{j,j+1} = 1$ and $\varphi_{j,j+1}=0$ for $j\in{\{1,...,N-2,N-1\}}$). We also assume no input drive phase $\varphi_\mathrm{in} = 0$ and no internal qubit decay or dephasing ($\Gamma'_{0,j} = 0$ and $\Gamma_{\varphi,j} = 0$ for $j \in \{1,...,N-1,N\}$). 

The transmission of the qubit chain is given by 

\begin{equation}
\label{eq:t chain}
t = 1 + \frac{\langle\hat{F}_N\rangle}{\alpha e^{i\varphi_\mathrm{in}} \prod_{j=1}^{N-1}\eta_{j,j+1} e^{i\varphi_{j,j+1}}}
\end{equation}

We use \cref{eq:t chain} to fit to the power-dependent transmission of $N$-qubit chains as shown in \cref{fig:Fig2}c,d. We also use the master equation to fit stabilized density matrices (\cref{appendix:Modeling:ddre fits}) and scattered radiation from the qubit chain (\cref{appendix:Modeling:photon sorting}). Simulations are performed using QuTiP \cite{qutip, qutip2}.

\subsection{Power-dependent cascaded qubit transmission}
\label{appendix:Modeling:power dep transmission}

The power-dependent cascaded qubit transmission shown in \cref{fig:Fig2} is fit to simulations of the master equation derived in \cref{appendix:Modeling:SLH ME}. Fitting parameters include emitter forward and backward waveguide decay rates, internal decay and dephasing, inter-qubit waveguide loss and phase, and input power.

For $N$ qubits, we observe splitting of the transmission profile into $N$ dips, as described in the main text. This is caused by hybridization between super-radiant and sub-radiant states in the single-photon manifold, which arises due to chiral waveguide-mediated interactions (see \cref{fig:Fig2}a). The transmission dip associated with each transition to a hybridized state becomes deeper with increased power (see \cref{fig:Fig2}c,d), which is consistent with power broadening for a strongly-coupled chiral qubit \cite{joshi2023resonance}. Further increasing the input power results in full saturation of the $N$-qubit chain, leading to unity transmission. Notably, collective qubit states in higher-excitation manifolds also hybridize, but do not play a significant role in measured transmission prior to full qubit saturation. For the highest experimental input powers, the total population in the single-excitation manifold is $\sim 0.02$ and the total excitation in higher-excited manifolds is $\sim 10^{-4}$.

\begin{figure*}[t!]
\centering
\includegraphics[width=1\linewidth]{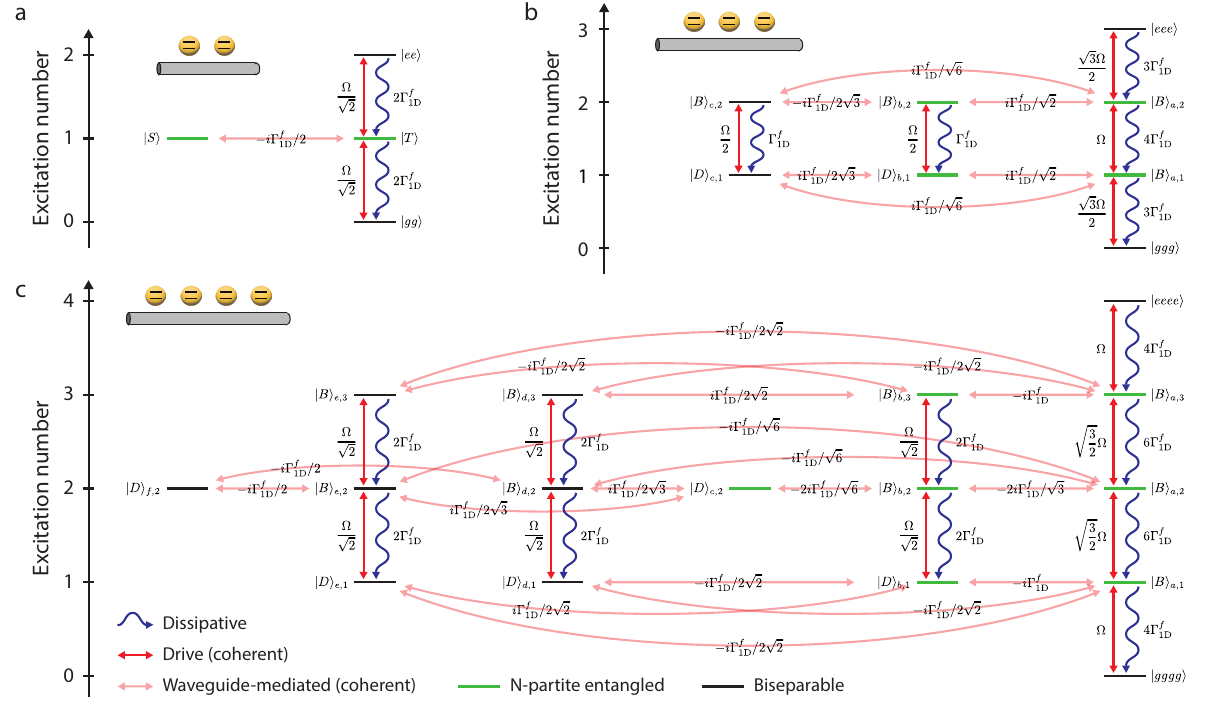}
\caption{\textbf{Level structures for $N$-qubit chains.} (a) Two qubits, where $|S\rangle= (|ge\rangle - |eg\rangle)/\sqrt{2}$ and $|T\rangle = (|ge\rangle+|eg\rangle)/\sqrt{2}$. (b) Three qubits. (c) Four qubits. Waveguide decay and coherent drives cause transitions within the same spin manifold, and particle-conserving waveguide-mediated interactions create coupling between states with the same excitation number.}
\label{fig:level structures}
\end{figure*}

For completion, we plot the level structures for $N=2$, $3$, and $4$ resonant cascaded qubits in \cref{fig:level structures}, obtained by addition of $N$ spin-1/2 particles. \cref{fig:level structures}a, b, and c show the two-, three-, and four-qubit cases, respectively. The three-qubit basis is chosen by Gram-Schmidt orthogonalization, and is given in \cref{tab:three-qubits}. The four-qubit basis is chosen by considering the analytical form of stabilized dark states given in \cite{pichler2015quantum}, and is presented in \cref{tab:four-qubits}. We also highlight the multipartite entangled states in \cref{fig:level structures} in green. 

\begin{table}[htpb]
\begin{ruledtabular}
\begin{tabular}{ll}
\textrm{Label} & \textrm{State} \\
\colrule
$|ggg\rangle$ & $|ggg\rangle$ \\
$|B\rangle_{a,1}$ & $\frac{1}{\sqrt{3}}\left( |egg\rangle + |geg\rangle+|gge\rangle \right)$ \\
$|B\rangle_{a,2}$ & $\frac{1}{\sqrt{3}}\left( |eeg\rangle + |ege\rangle+|gee\rangle \right)$ \\
$|eee\rangle$ & $|eee\rangle$ \\
\colrule
$|D\rangle_{b,1}$ & $\frac{1}{\sqrt{6}}\left( |egg\rangle + |geg\rangle - 2|gge\rangle \right)$ \\
$|B\rangle_{b,2}$ & $\frac{1}{\sqrt{6}}\left( 2|eeg\rangle - |ege\rangle - |gee\rangle \right)$ \\
\colrule
$|D\rangle_{c,1}$ & $\frac{1}{\sqrt{2}}\left(|egg\rangle-|geg\rangle \right)$ \\
$|B\rangle_{c,2}$ & $\frac{1}{\sqrt{2}}\left(|ege\rangle-|gee\rangle \right)$ \\
\end{tabular}
\end{ruledtabular}
\caption{Basis for three-qubit collective states.}
\label{tab:three-qubits}
\end{table}

\begin{table}[htpb]
\begin{ruledtabular}
\begin{tabular}{ll}
\textrm{Label} & \textrm{State} \\
\colrule
$|gggg\rangle$ & $|gggg\rangle$ \\
$|B\rangle_{a,1}$ & $\frac{1}{2}\left( |eggg\rangle + |gegg\rangle + |ggeg\rangle + |ggge\rangle \right)$ \\
$|B\rangle_{a,2}$ & $\frac{1}{\sqrt{6}} ( |eegg\rangle + |egeg\rangle+|egge\rangle +|geeg\rangle $ \\
 & $+|gege\rangle +|ggee\rangle)$ \\
$|B\rangle_{a,3}$ & $\frac{1}{2}\left( |eeeg\rangle + |eege\rangle + |egee\rangle + |geee\rangle \right)$ \\
$|eeee\rangle$ & $|eeee\rangle$ \\
\colrule
$|D\rangle_{b,1}$ & $\frac{1}{2} \left( |ggeg\rangle+|ggge\rangle-|gegg\rangle-|eggg\rangle \right)$ \\
$|B\rangle_{b,2}$ & $\frac{1}{\sqrt{2}} \left( |ggee\rangle-|eegg\rangle \right)$ \\
$|B\rangle_{b,3}$ & $\frac{1}{2} \left( |egee\rangle+|geee\rangle-|eeeg\rangle-|eege\rangle\right)$ \\
\colrule
$|D\rangle_{c,2}$ & $\frac{1}{2\sqrt{3}} ( 2|ggee\rangle + 2|eegg\rangle - |egeg\rangle - |gege\rangle $ \\
 & $- |egge\rangle - |geeg\rangle)$ \\
\colrule
$|D\rangle_{d,1}$ & $\frac{1}{\sqrt{2}} \left( |ggge\rangle-|ggeg\rangle \right)$ \\
$|B\rangle_{d,2}$ & $\frac{1}{2} \left( |egge\rangle + |gege\rangle - |egeg\rangle - |geeg\rangle \right)$ \\
$|B\rangle_{d,3}$ & $\frac{1}{\sqrt{2}} \left( |eege\rangle-|eeeg\rangle \right)$ \\
\colrule
$|D\rangle_{e,1}$ & $\frac{1}{\sqrt{2}} \left( |gegg\rangle-|eggg\rangle \right)$ \\
$|B\rangle_{e,2}$ & $\frac{1}{2} \left( |geeg\rangle + |gege\rangle - |egeg\rangle - |egge\rangle\right)$ \\
$|B\rangle_{e,3}$ & $\frac{1}{\sqrt{2}} \left( |geee\rangle-|egee\rangle \right)$ \\
\colrule
$|D\rangle_{f,2}$ & $\frac{1}{2} \left( |gege\rangle + |egeg\rangle - |egge\rangle - |geeg\rangle \right)$ \\
\end{tabular}
\end{ruledtabular}
\caption{Basis for four-qubit collective states.}
\label{tab:four-qubits}
\end{table}

\subsection{Driven-dissipative stabilization fits}
\label{appendix:Modeling:ddre fits}

We stabilize entangled four-qubit states using a waveguide drive, recovering full density matrices for variable-length drives (see \cref{appendix:Methods:state tomography}). This allows us to measure the relaxation dynamics of the driven qubit chain, which we fit to simulations of the master equation derived in \cref{appendix:Modeling:SLH ME}. Fits are applied with a loss function including the pair-wise concurrences, purity, and Von-Neumann entropy of the recovered density matrices as well as the measured qubit Pauli operators. Fitting parameters include emitter forward waveguide decay rates, inter-qubit phase, and input power. All other parameters are fixed using independently extracted values. We comment that the fitted parameters do not exactly match the values extracted in \cref{tab:table1} due to the large number of degrees of freedom in the model. Fit results are shown in \cref{fig:ddre}.

We compare the experimental steady-state density matrices with an ideal system driven at the same power. For the dimerized (multipartite) state, this corresponds to setting $\Gamma_{\mathrm{1D},j}^f=2$ MHz for all $j$ and $\delta_1 = 0 $ $(0)$ MHz, $\delta_2 = 0 $ $(2)$ MHz, $\delta_3 = 2 $ $(-2)$ MHz, and $\delta_4 = -2 $ $(0)$ MHz. The ideal case also has no intrinsic decoherence or inter-chip loss. For the dimerized (multipartite) state, we obtain $55.5^{+0.4}_{-0.7} \%$ $(62.8^{+0.6}_{-0.7}\%)$ fidelity between the experimental steady-state and the ideal state. We then estimate error budgets using master equation simulations fit to experimental data, which are given in \cref{tab:error}.

Between the two stabilized states, we observe the largest difference in the infidelity arising from intrinsic decoherence, with $26\%$ for the dimerized state and only $6\%$ for the multipartite state. This difference arises because we use a larger stabilizing drive power in the dimer case. It is a well known property of qubit-waveguide based stabilization protocols that larger pump powers result in increased entanglement generation at the cost of a larger relaxation time \cite{shah2024stabilizing}. In the presence of finite qubit decoherence, this increased stabilization time amplifies infidelity - as seen in the dimer. On the other hand, inter-chip loss more greatly affects the multipartite state (5$\%$ infidelity). The dimer state is nearly unaffected by inter-chip loss because it contains no inter-chip entanglement. In both cases, deviations from ideal detunings and waveguide decay rates contribute $\sim20\%$ to infidelity, which can be improved straightforwardly. 

\begin{table}[htpb]
\begin{ruledtabular}
\begin{tabular}{lll}
\textrm{ } & \textrm{Dimer} & \textrm{Multipartite} \\
\colrule
Internal decoherence & $26\%$ & $6\%$ \\
Inter-chip loss & $<1\%$ & $5\%$\\
Backward waveguide decay & $<1\%$ & $<1\%$\\
Ideal detunings, uniform $\Gamma_\mathrm{1D}^f$ & $18\%$ & $26\%$\\
\end{tabular}
\end{ruledtabular}
\caption{Error budget for entanglement stabilization}
\label{tab:error}
\end{table}

\subsection{Entanglement metrics}
\label{appendix:Modeling:ent}

We observe finite pair-wise concurrences for all qubit pairs of a four-qubit entangled state, shown in \cref{fig:ddre}c. However, this observation does not imply the presence of genuine multipartite entanglement. A quantum state containing three or more qubits is genuinely multipartite entangled if it is not biseparable - it cannot be written as a mixture of states that are separable across some bipartition. We detect the presence of genuine multipartite entanglement in our experiment using genuine multipartite negativity (GMN) \cite{jungnitsch2011taming}, which is a metric based on optimizing over potential entanglement witnesses. An entanglement witness is an observable which remains positive for biseparable states but takes on a negative value for at least one multipartite entangled state. The GMN is practically useful because it can be framed as a semi-definite program and computed numerically. It can also quantify multipartite entanglement. GMN takes on values ranging from 0-0.5, with GMN $\neq0$ indicating the presence of genuine multipartite entanglement. For maximally entangled states, like the GHZ state, GMN = 0.5. We compute the GMN for the stabilized state of \cref{fig:ddre}c following \cite{jungnitsch2011taming}, obtaining GMN = $6.3^{+0.8}_{-0.5}\times10^{-2}$ (95$\%$ confidence interval). We note that for ideal settings with infinite Purcell factors and strong drives, GMN $= 0.5$ should be achievable. 

\subsection{Photon bound state fits}
\label{appendix:Modeling:photon sorting}

We compare measured data of photon bound states to master equation simulations. This requires obtaining the quantum state of the field scattered from a pulse incident on the cascaded qubits. Typically, master equation approaches trace out the bath modes and focus on the qubit states. In our case, because the waveguide modes contain the relevant information and vary with time, we resort to performing pulsed input-output simulations following \cite{pulsed_inputoutput_molmer}. In this formalism, the incoming pulse is modeled as a cavity $\hat{a}_{u}$ upstream to the qubit chain, with a time-dependent coupling to the waveguide to account for the pulse shape. If the shape of the input pulse is given by $u(t)$, the required time-dependent coupling is given by

\begin{align}
    g_u(t) = \frac{u^*(t)}{\sqrt{1 - \int_0^t dt' |u(t')|^2}}
\end{align}

In a similar fashion, this formalism can be used to determine the quantum state of a given mode at the output. In our case, we are simply interested in the correlations as a function of time, rather than choosing specific modes. Hence, we obtain the outgoing field with:
\begin{align}
    \hat{L}_{0}(t) = \sum_{j}\sqrt{\Gamma_{\mathrm{1D}, j}^{f}}\hat{\sigma}_{j}^{-} + g^{*}_{u}(t)\hat{a}_{u}
\end{align}
where $\hat{\sigma}_{j}^{-}$ is the lowering operator for qubit $j$. This is equivalent to the emitted field in the absence of loss in the waveguide and backward emission. We account for waveguide loss following \cite{reuer2022realization}. The same-time correlators are then found using $G^{(j)}(t, t) = \langle [\hat{L}_{0}^{\dagger}(t)]^{j}[\hat{L}_{0}(t)]^{j} \rangle$. \par

Simulation parameters are obtained by fitting to the experimental data, taking separately extracted qubit parameters as initial values. The only fitted parameters are the emitter forward waveguide decay rates. All other experimental parameters are kept fixed. Similar to \cref{appendix:Modeling:ddre fits}, we note that fitted parameters do not exactly match values obtained in \cref{tab:table1}. The delay for the $n$ photon wavepacket, $\tau_{n}$, is analytically derived by using the delay for each emitter as $4/(\Gamma_\mathrm{1D}^f n^{2})$ where $\Gamma_\mathrm{1D}^f$ is the fitted forward decay rate of the corresponding emitter. \par

\end{document}